\documentclass[sigconf]{acmart} 

\usepackage{makecell}
\usepackage{array}
\usepackage{multirow}
\usepackage{makecell}
\usepackage{enumitem}
\usepackage{xcolor}
\usepackage{balance}

\definecolor{mybg}{HTML}{F2F9FD}
\definecolor{myframe}{HTML}{7DA0B0}

\AtBeginDocument{%
  }

\setcopyright{acmlicensed}
\copyrightyear{2026}
\acmYear{2026}
\setcopyright{cc}
\setcctype{by}
\acmConference[CHI '26]{Proceedings of the 2026 CHI Conference on Human Factors in Computing Systems}{April 13--17, 2026}{Barcelona, Spain}
\acmBooktitle{Proceedings of the 2026 CHI Conference on Human Factors in Computing Systems (CHI '26), April 13--17, 2026, Barcelona, Spain}
\acmPrice{}
\acmDOI{10.1145/3772318.3791244}
\acmISBN{979-8-4007-2278-3/2026/04}

\acmSubmissionID{2520}

\newcommand{\add}[1]{\textcolor{black}{#1}}
\begin{document}

\title[Redundant is Not Redundant: Automating Efficient Categorical Palette Design]{\textit{Redundant is Not Redundant}: Automating Efficient Categorical Palette Design Unifying Color \& Shape Encodings with \textit{CatPAW}}

\author{Chin Tseng}
\affiliation{%
   \institution{University of North Carolina at Chapel Hill}
   \city{Chapel Hill}
   \state{NC}
   \country{USA}}
\email{chint@cs.unc.edu}

\author{Arran Zeyu Wang}
\affiliation{%
   \institution{University of North Carolina at Chapel Hill}
   \city{Chapel Hill}
   \state{NC}
   \country{USA}}
\email{zeyuwang@cs.unc.edu}

\author{Ghulam Jilani Quadri}
\affiliation{%
   \institution{University of Oklahoma}
   \city{Oklahoma City}
   \state{OK}
   \country{USA}}
\email{quadri@ou.edu}

\author{Danielle Albers Szafir}
\affiliation{%
   \institution{University of North Carolina at Chapel Hill}
   \city{Chapel Hill}
   \state{NC}
   \country{USA}}
\email{danielle.szafir@cs.unc.edu}

\begin{teaserfigure}
  \centering
  \includegraphics[width=\linewidth]{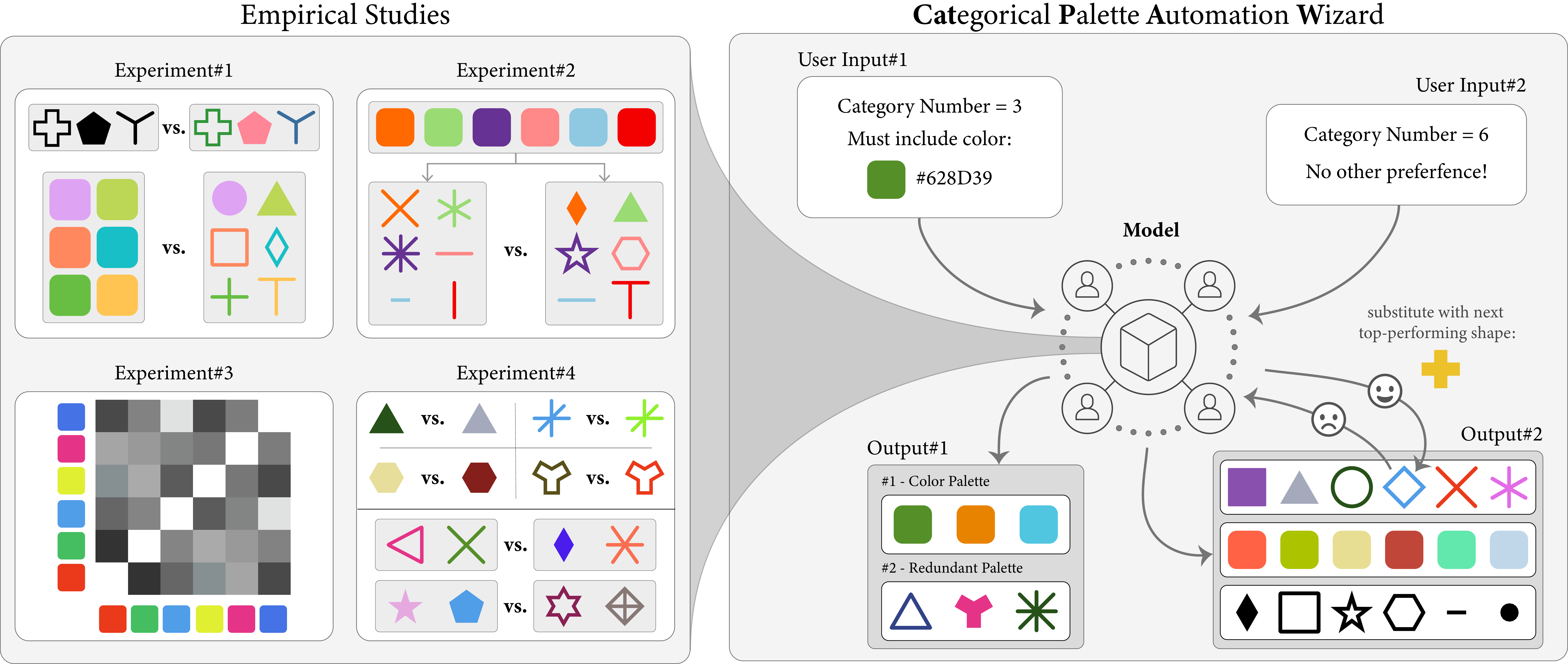}
  \caption{We present \href{https://catpaw-categorical-palette.web.app/}{\emph{CatPAW}}, a web-based tool for generating effective categorical palettes using a data-driven model built from four crowd-sourced experiments. Users can specify palette requirements, such as the number of categories, must-include colors or shapes, and palette type (color-only, shape-only, or redundant). Based on performance estimates drawn from empirical data, \emph{CatPAW} suggests a set of optimal palettes adhering to the provided constraints. Users can interactively swap out undesired elements, and \emph{CatPAW} will suggest high-ranking alternatives by considering the current palette composition. Rounded rectangles represent color-only palettes.}
  \label{fig:teaser}
\end{teaserfigure}

\begin{abstract}
Colors and shapes are commonly used to encode categories in multi-class scatterplots. Designers often combine the two channels to create redundant encodings, aiming to enhance class distinctions. However, evidence for the effectiveness of redundancy remains conflicted, and guidelines for constructing effective combinations are limited. This paper presents four crowdsourced experiments evaluating redundant color–shape encodings and identifying high-performing configurations across different category numbers. Results show that redundancy significantly improves accuracy in assessing class-level correlations, with the strongest benefits for 5–8 categories. We also find pronounced interaction effects between colors and shapes, underscoring the need for careful pairing in designing redundant encodings. Drawing on these findings, we introduce a categorical palette design tool that enables designers to construct empirically grounded palettes for effective categorical visualization. Our work advances understanding of categorical perception in data visualization by systematically identifying effective redundant color–shape combinations and embedding these insights into a practical palette design tool.
\end{abstract}




\maketitle

\section{Introduction}

Effective approaches for representing categorical data are crucial for data visualization ~\cite{szafir2023visualization}; however, we have limited guidance as to how to effectively encode categorical data~\cite{tseng2024shape,graze2024building}. 
Categorical encodings principally use color or shape \textit{palettes}, predefined collections of shapes or colors that are then assigned to different categories~\cite{ware2012information}.
Designers frequently 
differentiate categories using both color and shape simultaneously to increase each category's perceptual distinctiveness.
Such \textit{redundant encoding} is assumed to increase people's abilities to distinguish between different categories %
and has been widely applied in real-world scenarios from scientific papers~\cite{wang2023human} to business reports~\cite{charumilind2022will}.
However, the benefits of redundant encoding are, in part, contentious, with some studies showing little to no impact on effectiveness \cite{gleicher2013perception}, others showing notable benefits for given tasks \cite{nothelfer2017redundant}, and still others showing that the channels themselves may interact, suggesting more complex effects for redundancy \cite{smart2019measuring}. 
Part of this tension may arise from the lack of empirically-grounded guidance for designing effective categorical visualization~\cite{wang2025characterizing}: categorical palettes are largely designed based on heuristics, leading to potential inconsistencies in encoding effectiveness.
Existing guidance for redundant palette design treats the two channels individually, either helping designers build a color palette~\cite{gramazio2016colorgorical} or a shape palette~\cite{tseng2024shape} and assuming their pairing will lead to improved visualizations. We lack clear guidance for how to effectively pair color and shape to maximize the potential benefits of redundancy.

In this paper, we explore redundant encoding design to understand its utility and derive empirical guidance for effective redundant palette design. 
Part of the challenge in reconciling color and shape for palette design is that the two spaces are computationally incompatible. 
Color is sampled from continuous numerical spaces; therefore, it can fit into a diverse range of design and creation processes.
For example, individual color palettes can be created based on manual selection~\cite{harrower2003colorbrewer}, mathematical color metrics~\cite{gramazio2016colorgorical}, or even machine learning models~\cite{hong2024cieran} sampling over continuous spaces.
However, shape lacks a common space or set of agreed-upon features and is less likely to have a universal descriptive language \cite{burlinson2017open, ware2012information}. Even coarse-grained features describing shape fail to explain shape palette effectiveness \cite{tseng2024shape}. 
Past approaches for constructing shape palettes rely on preselected shapes or on empirical data used to build pairwise accuracy models~\cite{tseng2024shape}.
Therefore, even though the use of redundant encoding can improve 
accuracy for tasks like visual segmentation and grouping~\cite{nothelfer2017redundant}, %
we still lack an understanding of how to effectively design redundant palettes and how specific redundant encodings (i.e., color-shape pairings) affect task performance in data visualization~\cite{wang2025characterizing}.

We address this knowledge gap using a data-driven approach, collecting
empirical data about expert-crafted palettes to generate a method for crafting effective redundant palettes. 
We developed this method in a series of four crowdsourced experiments asking people to compare correlations across different categories. We structure these experiments to first evaluate the effectiveness of redundant encodings and then establish effective color-shape pairings in multi-class scatterplots across varying numbers of categories. Our studies first investigate the effectiveness of redundancy for designer palettes (Experiment 1) and then explore the interactions of shape and color in redundant palette design (Experiment 2). 
Finally, we construct pairwise accuracy matrices for colors in redundant mappings (Experiment 3) and for subsequent interactions between colors and shapes (Experiment 4) to develop a statistical model measuring the performance of color-shape pairings 
to guide effective redundant palette design. 

Our findings indicate that redundant encodings improve accuracy in assessing class-level correlations, in line with both heuristic guidance and past studies \cite{nothelfer2017redundant}. 
Redundancy best
benefited performance 
in visualizations with moderate category numbers (five to eight categories). 
Furthermore, 
performance varied between different combinations of color and shape palettes, emphasizing the importance of choosing appropriate redundant encoding combinations. Notably, effective redundant palette design is not as simple as blending high-performing color and shape palettes: combining top-performing shapes and colors did not consistently result in top-performing redundant encodings.

Using the data from our studies, we developed a statistical model that generates effective redundant palettes based on the number of categories present in a visualization.
We incorporate this model into a design tool, CatPAW—\emph{Categorical Palette Automation Wizard}—which assists designers in creating optimized categorical palettes. CatPAW supports color-only, shape-only, and color–shape redundant encodings tailored to different category numbers and user preferences for color or shape. By unifying the color and shape channels, the tool facilitates more informed and efficient palette design.
Overall, this work contributes to a more integrated understanding of redundant encoding for categorical perception in data visualization and offers an actionable framework for encoding categorical data.

\noindent \textbf{Contribution:}
We empirically investigated the effectiveness of redundant encodings, specifically color and shape pairings, in data visualizations. Our findings help reconcile contradictory evidence on the effectiveness of redundancy, showing that redundancy can benefit categorical analysis, but that this benefit is in part dependent on the number of categories as well as the palette design.
Our results built up a deeper understanding of how redundant encodings improve people's perception and overall task performance in visualizations and suggest that optimal redundant palette design should consider the interplay between color and shape.
Driven by those insights, we created an automated palette generation tool %
that leverages the data generated in our experiments to aid designers in creating categorical encodings.

\section{Related Work}

Providing empirical evidence for design heuristics is crucial for effective visualization~\cite{szafir2023visualization, elliott2020design, wang2025characterizing, kosara2016empire}.
In this section, we review 
prior research 
on the use of redundant encodings in data visualization, focusing
on three key areas: graphical perception,
the role of color and shape in categorical perception, and 
redundant encoding in data visualization.

\subsection{Graphical Perception of Scatterplots}

Graphical perception refers to people's ability to interpret and understand information presented in graphical form~\cite{szafir2023visualization, munzner2014visualization}.
Graphical perception studies explore how visual encoding and design elements in visualizations, such as position~\cite{hong2021weighted}, axis~\cite{long2024cut}, angle~\cite{wang2017there}, and color~\cite{schloss2018mapping}, are perceived and processed by people to provide empirical guidance for visualization design.
Starting from Cleveland and McGill's initial work four decades ago~\cite{cleveland1984graphical}, prior research on graphical perception identified a hierarchy of perceptual effectiveness across different tasks, demonstrating that certain visual encodings are more effective than others for conveying quantitative information~\cite{quadri2021survey}.

Scatterplots are among the most widely used visualization types in graphical perception research \cite{rensink2013prospects}. Designers often use scatterplots to encode multiple categories of data in the same space, enabling direct comparisons between different categories of data
\cite{munzner2014visualization, szafir2023visualization, sarikaya2017scatterplots}. 
As a result, scatterplots have 
been used to derive a variety of insights on perceptual efficiency in visualization design, both for improved scatterplots and for general data representation practices~\cite{rensink2013prospects, rensink2013prospects, quadri2021survey}.
Graphical perception experiments have used scatterplots to explore effective visualization design in a diverse range of tasks, including value judgments~\cite{gleicher2013perception}, correlation~\cite{harrison2014ranking}, comparison~\cite{gleicher2011visual}, trend~\cite{correll2017regression}, clustering~\cite{sedlmair2012taxonomy}, and uncertainty~\cite{sarma2022evaluating}, and visual encodings, such as color~\cite{mayorga2013splatterplots}, shape~\cite{burlinson2017open}, aggregations~\cite{xiong2019illusion}, and texture~\cite{bachthaler2008continuous}.
Among different types of scatterplots, %
multiclass scatterplots, our focus in this work, have played a pivotal role in understanding the visual data communication process~\cite{franconeri2021science}.

\subsection{Color and Shape in Categorical Perception}

Categorical perception involves the brain's ability to categorize continuous stimuli into discrete groups~\cite{goldstone2010categorical}.
In the context of data visualization, 
we take advantage of these processes to represent different classes of data by applying encodings that intuitively create distinct visual groups~\cite{demiralp2014learning,tseng2023evaluating, tseng2024revisiting}.
Color and shape are the channels most commonly used to differentiate between categories~\cite{szafir2023visualization, wang2025characterizing}.
People can efficiently process categorical differences in color~\cite{tseng2023evaluating} and shape~\cite{tseng2024shape}, but the effectiveness of these encodings can vary depending on several factors~\cite{wang2025characterizing}. 

Research in cognition and vision science 
describes how different color attributes, such as hue~\cite{borland2007rainbow}, saturation~\cite{zhou2015survey}, and lightness~\cite{schloss2010aesthetics}, affect our ability to perceive and interpret categorical information.
Hue is often considered a primary color feature to consider for categorical data, with best practices emphasizing mapping classes to distinct hues~\cite{harrower2003colorbrewer,graze2024building,stone2006choosing}.
Predominantly hue-varying categorical color palettes 
can achieve high levels of accuracy in tasks like comparing category means
\cite{tseng2023evaluating}.
However, hue in isolation is often insufficient for effective visualization design. 
Lightness and saturation also contribute to 
people's abilities to distinguish categorical data~\cite{tseng2024revisiting,samsel2018art}.
Beyond the fundamental perceptual attributes of color, its psychological and cultural associations can also influence how viewers interpret categorical data visualizations.
Color 
carries semantic or affective meaning that may be more important than common perceptual factors for visualization~\cite{schloss2018mapping, zimnicki2023effects,lin2013selecting,bartram2017affective}.
For example, warm colors like red and orange might be associated with urgency or excitement~\cite{kinateder2019color}, while cool colors such as blue and green may convey calmness or stability~\cite{schloss2020blue}.
Although not the central focus of this study, these factors can influence user interpretation, preference, and overall engagement with the data; however, their utility and design typically must align with the semantic content of the data or its associated context.

Shape is another
fundamental visual channel for categorical data encoding
\cite{ware2012information}.
Shape is considered an effective preattentive attribute~\cite{huang2020space}, meaning that it can be processed by the visual system quickly and without conscious effort, making it particularly useful for conveying categorical information at a glance.
The perceptual discriminability of shapes is crucial for 
effective categorical encoding; however, we lack formalized computational spaces for reasoning about shape.
Past efforts have categorized widely-used shapes in data visualization using broad categories, such as open and closed~\cite {burlinson2017open}. Q-Tons~\cite{ware2009quantitative} offer a linear sequence of shapes for ordinal data, but, like most shape palettes, are hand-designed to capture the idea of an "increasing" shape value.
Despite the lack of standard spaces for reasoning about shape, shape remains a common choice for visualization, often preferred to color due to its increased accessibility for people with color vision deficiency (CVD)~\cite{stone2006choosing}.
Guidelines for selecting shape palettes for data visualization are limited, but suggest considering the distinctiveness of shape and ease of visual processing~\cite{demiralp2014learning}. Existing shape palette design relies heavily on intuition and the designer's preferences in real-world practices.
More recently, \emph{Shape It Up} explored a data-driven approach 
to create effective shape palettes based on experimental data collected over a series of 
categorical data analysis tasks~\cite{tseng2024shape} but focused exclusively on monochromatic shapes.

Past research into both shape and color perception for visualization usually considers each channel individually~\cite{wang2025characterizing}.
While limited past evidence shows the two channels may interact \cite{smart2019measuring}, these studies focus on analyzing small differences in color better suited to continuous data. Despite their common pairing~\cite{wang2023human}, 
we lack an understanding of how these channels may interact in categorical encodings. 
In this work, we aimed to combine color and shape together to measure their performance, better understand the influence of redundancy on categorical encoding, and build more efficient palette design tools.

\subsection{Redundant Encoding in Data Visualization}

Redundant encoding
refers to the practice of using more than one visual variable to represent the same underlying data.
A widely-used redundant encoding design combines color and shape. For instance, one category might be represented by red circles, while another is shown as blue squares.
People employ redundant encodings largely to make it easier to distinguish between different categories of data \cite{nothelfer2017redundant}. 
Redundant encoding with color and shape can also offer improved accessibility, particularly for individuals with CVD~\cite{barstow2019examining}.
For viewers who have difficulty distinguishing between certain colors, the shape of the data point can serve as an alternative visual cue, allowing them to still effectively differentiate between categories~\cite{iso9186}.
Conversely, for individuals with typical color vision, color can reinforce the information shape conveys, potentially leading to faster and more accurate perception~\cite{iso9186}.
However, employing redundancy means that there are fewer visual channels available to encode data and can increase the complexity of the visualization. 

Redundant encoding, especially for scatterplots, is available in a number of commercial and research toolkits. 
For example, 
Tableau~\cite{tableau}, R~\cite{R}, and Matlab~\cite{MATLAB} all allow people to combine colors and shapes in one encoding; however, these combinations reflect each tool's default shape and color palettes and require manual adjustment by the user to change. 
The \textit{scatterHatch} package in the R/Bioconductor~\cite{guha2022generating} is specifically designed to create plots for massive single-cell data that use redundant coding by combining colors with patterns.
This package aims to enhance the accessibility and distinguishability of large-scale single-cell spatial data in bioinformatics, which typically contain large numbers of categories.
However, scatterHatch primarily focuses on bioinformatics data, and the encoded patterns are usually not exactly shapes but their variations, such as a textured plot with obscured dots, which is not commonly used in real-world data communication~\cite{franconeri2021science, szafir2023visualization, borkin2013makes}.

Despite its popularity, redundant encoding remains controversial. 
Previous research has demonstrated that redundant encodings can improve the speed and accuracy of typical vision science tasks, including data segmentation and grouping~\cite{nothelfer2017redundant}.
However, alternative studies have found little to no benefit for redundant shape and color encoding in aggregation tasks in visualizations \cite{gleicher2013perception}. 
Detractors also argue that redundant encoding comes at a significant cost.
For instance, encoding too much or irrelevant information can clutter the visualization and hinder rather than help the viewer's understanding~\cite{ellis2007taxonomy}.
In the context of scatterplots, which naturally require higher cognitive processing~\cite{cleveland1984graphical, quadri2024do}, using too many visual dimensions simultaneously, even if intended as redundant encodings, may make the graph appear more complex and more difficult to read.
Shape and color may interact perceptually. That is, shape may change perceptions of color and color of shape~\cite{smart2019measuring}, making it harder to merge these two channels without %
better models of their interaction. However, we lack informative guidance for crafting redundant encodings; instead, most popular tools typically pair the user's chosen color and shape palettes without regard to the interactions of the associated colors and shapes. 

Our research 
builds upon the existing tensions surrounding redundant encodings 
to better characterize effective redundant encoding design. We achieve this by systematically investigating specific color-shape redundant pairings 
across a range of conditions. We begin by confirming the effectiveness of redundancy and then develop a series of pair-wise matrices describing color-, shape-, and combined color-shape effectiveness in redundant encoding design using empirical data. We use these matrices to 
design a tool to help people create effective redundant palettes.

\section{Experiment One: Do Redundant Encodings Help?}
\label{sec-exp1}

Our first study examined how the choice of non-redundant encodings (color-only or shape-only) versus redundant encodings (color-and-shape) impacts people's abilities to reason across multiclass scatterplots. We performed a crowdsourced study measuring people's performance in comparing category correlations over varying category numbers ($N=2$--$10$).

We hypothesized that:
\noindent\textbf{H$_{1}$:} \textbf{(a) Redundant encodings would improve overall performance,} and \textbf{(b) differences in accuracy between redundant and non-redundant encodings will be larger as category number grows, but may plateau at a certain category number. }
Redundant encodings use an additional channel to encode categorical data. We expect people may benefit from having an extra reference for differentiating categories \cite{ware2012information,nothelfer2017redundant}.
People may be relatively strong at distinguishing between categories for smaller numbers of categories~\cite{wang2025characterizing}, such as the three category sets in Gleicher et al.\cite{gleicher2013perception}; however, we expect that the benefits of added perceptual cues emerge as the number of categories (and, as a result, difficulty of categorical analysis) increases.
As performance tends to be asymptotic with respect to the number of categories in single encodings~\cite{wang2025characterizing}, we anticipate the benefits of redundant encodings will likewise eventually saturate as the visualizations increase in complexity~\cite{tseng2023evaluating, tseng2024shape}.

\subsection{Experimental Design}
\subsubsection{Task}
We employed a correlation judgment task, which is one of the most widely employed tasks in previous studies of multiclass data visualization~\cite{harrison2014ranking, kay2015beyond, rensink2010perception} and is frequently employed in a range of real-world applications, including data reporting~\cite{ritchie2019age} and journalism~\cite{badger2018s}.
The task asked participants to identify the category in a multiclass scatterplot with the highest correlation.
While this task likely engages similar perceptual mechanisms to other aggregation tasks like averaging \cite{szafir2016four,tseng2023evaluating}, it represents another common use of scatterplots where estimation accuracy depends on people's ability to differentiate between categories \cite{sarikaya2017scatterplots}.

\begin{figure}[htbp] 
\centering
\vspace{0em}
\includegraphics[width=0.45\textwidth]{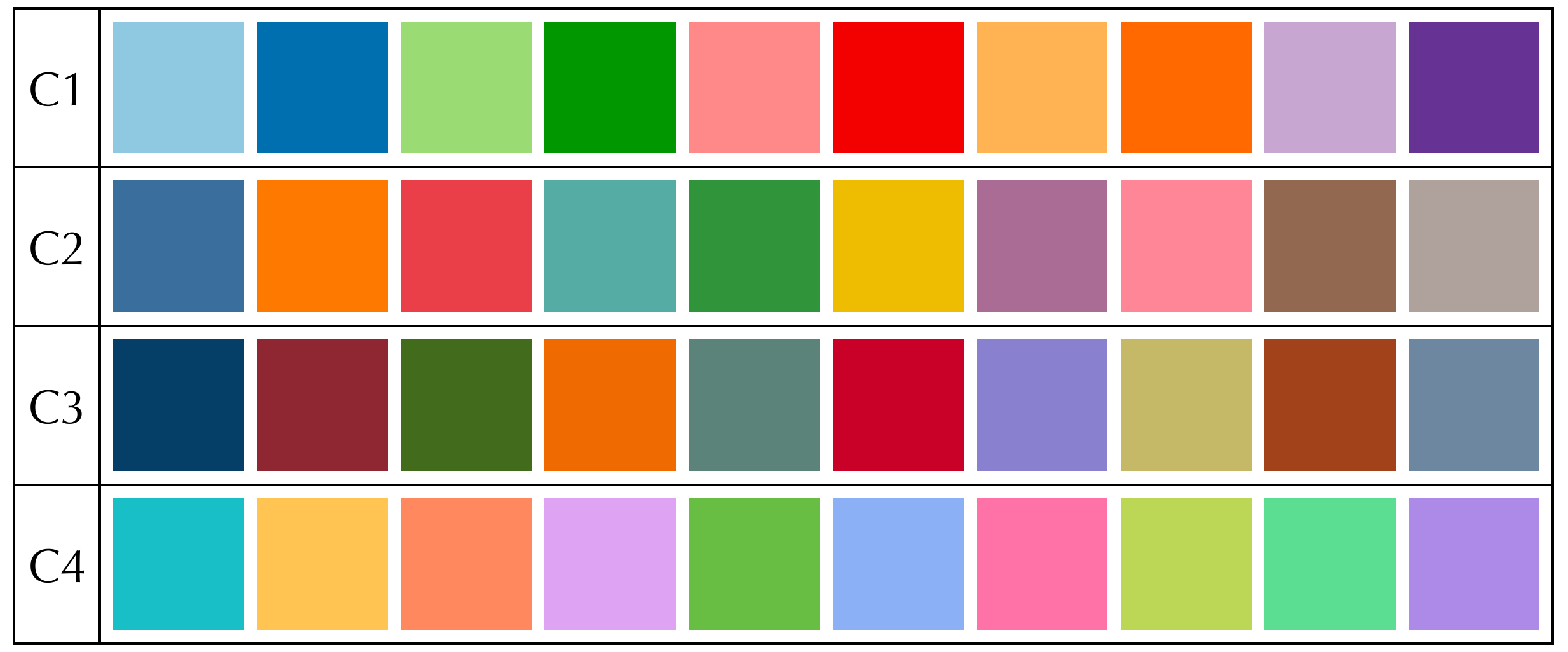} 
\vspace{0em}
\caption{The four color palettes used in Experiment 1 and 2. Each palette has 10 colors, drawn from (1) ColorBrewer/Paired~\cite{harrower2003colorbrewer}, (2) Tableau/Tab10~\cite{tableau}, (3) Stata/S2~\cite{statagraphics19}, and (4) Carto/Pastel~\cite{carto}}
\label{fig:color-palette}
\end{figure}

\subsubsection{Stimulus Generation}
\begin{figure}[htbp] 
\centering
\includegraphics[width=0.45\textwidth]{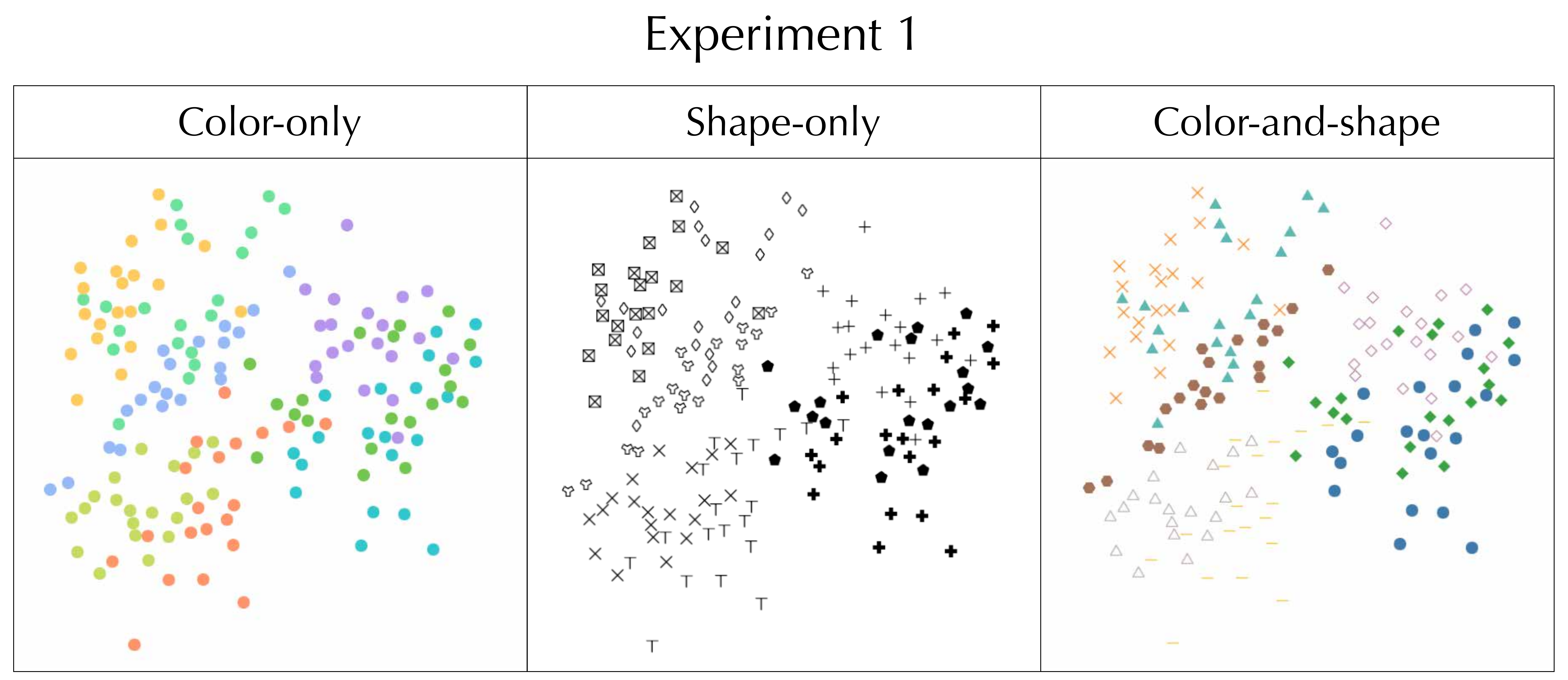} 
\caption{Example stimuli from Experiment 1: scatterplots encoding categorical data using color-only, shape-only, and redundant (color-and-shape) encodings. All three visualizations use the same data distribution with 8 categories. Correlation comparison tasks were used in Experiment 1.} 
\vspace{-1em}
\label{fig:stimuli_exp1}
\end{figure}
Participants estimated category correlations across a series of scatterplots using either colors, shapes, or color-and-shape to differentiate categories. We generated each scatterplot as a 400$\times$400 pixel graph using D3, as shown in \autoref{fig:stimuli_exp1}. Each scatterplot was rendered to a white background and two orthogonal black axes with 13 unlabeled ticks. 

We used four color palettes found in existing tools, shown in \autoref{fig:color-palette}, to encode categorical data in scatterplots: 1. ColorBrewer/Paired~\cite{harrower2003colorbrewer}, 2. Tableau~\cite{tableau}, 3. Stata/S2~\cite{statagraphics19}, and 4. Carto/Pastel~\cite{carto}. We chose these color palettes from previous studies~\cite{tseng2023evaluating} and based on their differences in $L^*$ variance, $L^*$ magnitude, and all-pairs perceptual distance~\cite{sharma2005ciede2000}, which are significant factors that may impact categorical perception. Such selection can also reduce bias from different color palettes. 

We encoded shape using 39 shapes collected from a previous study \cite{tseng2024shape}. The 39 shapes were derived from palettes in popular visualization tools: D3~\cite{6064996}, Tableau~\cite{tableau}, Excel~\cite{msexcel}, R~\cite{R}, and Matlab~\cite{MATLAB}. For redundant encodings, we preconstructed palettes by randomly selecting both colors and shapes from the above collections for each trial. For instance, in a scatterplot with five categories, five colors were randomly chosen from a specific color palette, while five shapes were randomly selected from the set of 39. These colors and shapes were then randomly paired, with each color assigned to a single shape, and each combination represented a unique category. The randomization helps minimize bias related to the ordering of colors and shapes or their pairings.

Points were rendered in a 6$\times$6-pixel window. 
Point locations were generated using NumPy’s random multivariate method~\cite{2020NumPy-Array}, ensuring that the target category had the highest Pearson correlation coefficient (ranging from 0.8 to 0.95), while the second-highest category maintained a correlation at least 0.2 lower than the target, consistent with thresholds used in past studies \cite{tseng2024shape}. 
All other categories had a randomly-selected correlation below this threshold. Each category consistently contained 20 points. 
To minimize confounding effects from spatial overlap, we applied jittering to any overlapping points. This point generation strategy was applied consistently across all experiments.

To evaluate whether redundant encodings enhance performance on multiclass scatterplots across varying numbers of categories, 
we generated datasets containing $N = [2-10]$ categories. 
\add{We use an upper bound of 10 categories as this aligns with recent studies of categorical encoding \cite{tseng2023evaluating,tseng2024shape}, reflects the size of the smallest tested color palette (Tableau 10), allows us to keep the study tractable in terms of the number of tested conditions, and, in piloting, we found that larger category numbers exhibited little change in performance, consistent with past findings on Fetchner's Law in categorical scatterplots \cite{wang2025characterizing}.}
As an exhaustive comparison of all colors and shapes in our encoding pool is prohibitively large, for each category number, we generated 20 different  sets \add{for each encoding}: \add{20 color palettes, each consisting of five sets of N random colors 
selected from each of the four color palettes,} and 20 shape sets, \add{each generated by randomly selecting N different shapes} from the pool of 39 shapes. We paired these two sets of colors and shapes to create the set of redundant encodings. These resulting 540 stimulus designs (3 encoding types $\times$ 20 encoding sets $\times$ 9 category numbers), which were divided into 10 task groups. The category numbers and encoding type were equally distributed across task groups, with each containing 54 stimuli.

\subsubsection{Procedure}
Our experiment was structured into three phases: (1) obtaining informed consent, (2) task description and tutorial, and (3) the formal study. Initially, participants provided informed consent in accordance with our IRB protocol and then supplied demographic information.  Next, they were introduced to the correlation task description alongside example scatterplots demonstrating varying levels of correlation. \add{Examples of the instructional and study interfaces are shown in the \href{https://osf.io/4up7j/?view_only=43e5681b71fc4fdb92617b8cef27c6c0}{OSF Supplements}.}
Participants were asked to complete a tutorial with three practice questions, each with a different encoding type (color, shape, and redundant encoding). They were required to answer all tutorial questions correctly to ensure task understanding before 
advancing. 

During the formal study, participants performed our target task (``Identify which category is the most correlated'') for 57 stimuli (54 formal trials and 3 engagement checks) presented sequentially
in a randomized order. At the start of the study, each participant was randomly assigned to one of 10 task groups.
Participants had 20 seconds to respond to each stimulus, after which time the answer was marked as incorrect and the study advanced to the next trial. To ensure valid participation, we included three engagement checks. These engagement checks were stimuli with two or three categories that had at least a $\Delta r = 0.4$ difference in their Pearson correlation coefficient \add{(see the \href{https://osf.io/4up7j/?view_only=43e5681b71fc4fdb92617b8cef27c6c0}{OSF Supplements} for examples)}. We randomly placed the engagement checks throughout the 54 formal trials.

\subsubsection{Participants}
We recruited 105 participants through Amazon Mechanical Turk (MTurk), requiring a minimum 95\% approval rating and residency in the US or Canada. Five participants who failed more than one engagement check were excluded from the analysis. The final data included 100 participants (72 male, 28 female), ranging in age from 20 to 65 years, with 10 participants assigned to each task group. All participants reported having normal or corrected-to-normal vision. \add{Our experiment took 10 minutes on average, and each participant was compensated \$1.60 for their time.}

Crowdsourcing can introduce inherent variation in stimulus appearance. This variation is ecologically consistent with visualization use and aligns with a wealth of past graphical perception research. These trade-offs limit our abilities to make precise claims about human vision, but do enable us to draw conclusions about graphical perception appropriate to the context of visualization design and use. We follow best practices for crowdsourced study design \cite{borgo2018information, heer2010crowdsourcing}. To maintain consistent stimulus sizes, we restricted participation to devices no smaller than standard laptops or monitors. We additionally leveraged colors that were robustly different in crowdsourced environments at our tested stimulus size \cite{szafir2018modeling}.

\subsubsection{Analysis}
We used accuracy as our primary dependent measure. To compare the performance between non-redundant encodings (color-only or shape-only) and redundant encodings (color-and-shape), we analyzed the resulting data using an ANOVA with category number and encoding type as independent variables. \autoref{tab:exp1} summarizes our results. The anonymized data and results for our study can be found on \href{https://osf.io/4up7j/?view_only=43e5681b71fc4fdb92617b8cef27c6c0}{OSF Supplements\footnote{\add{\url{https://osf.io/4up7j/?view_only=43e5681b71fc4fdb92617b8cef27c6c0}}}}.

\begin{table}[t] 
\centering
\caption{ANOVA results for category number and encode type (color-only, shape-only, color-and-shape). Significant effects are indicated by \textbf{bold} text.}
\includegraphics[width=1\linewidth]{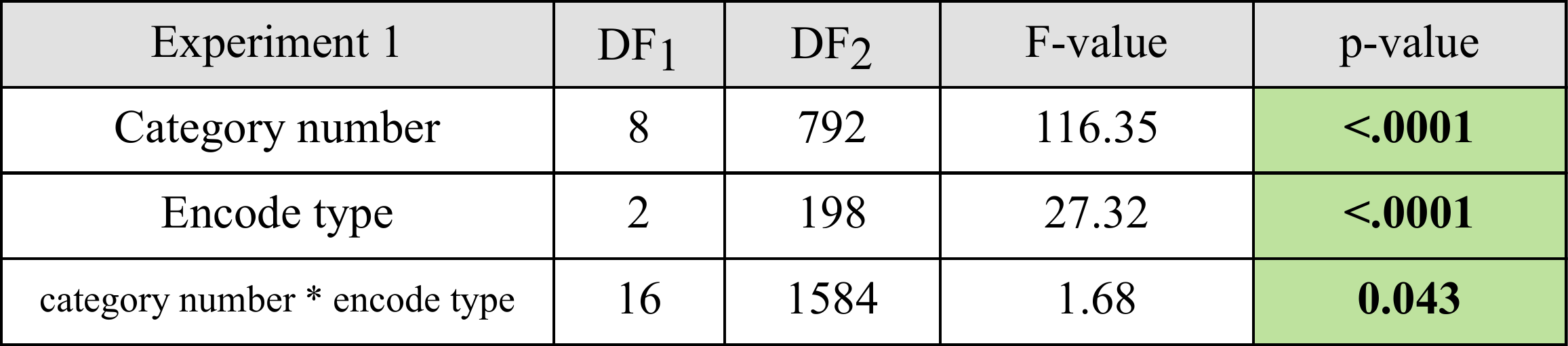} 
\vspace{-0.5em}
\label{tab:exp1}
\end{table}

\subsection{Results}
\begin{figure}[t] 
\centering
\vspace{-0.2em}
\includegraphics[width=0.45\textwidth]{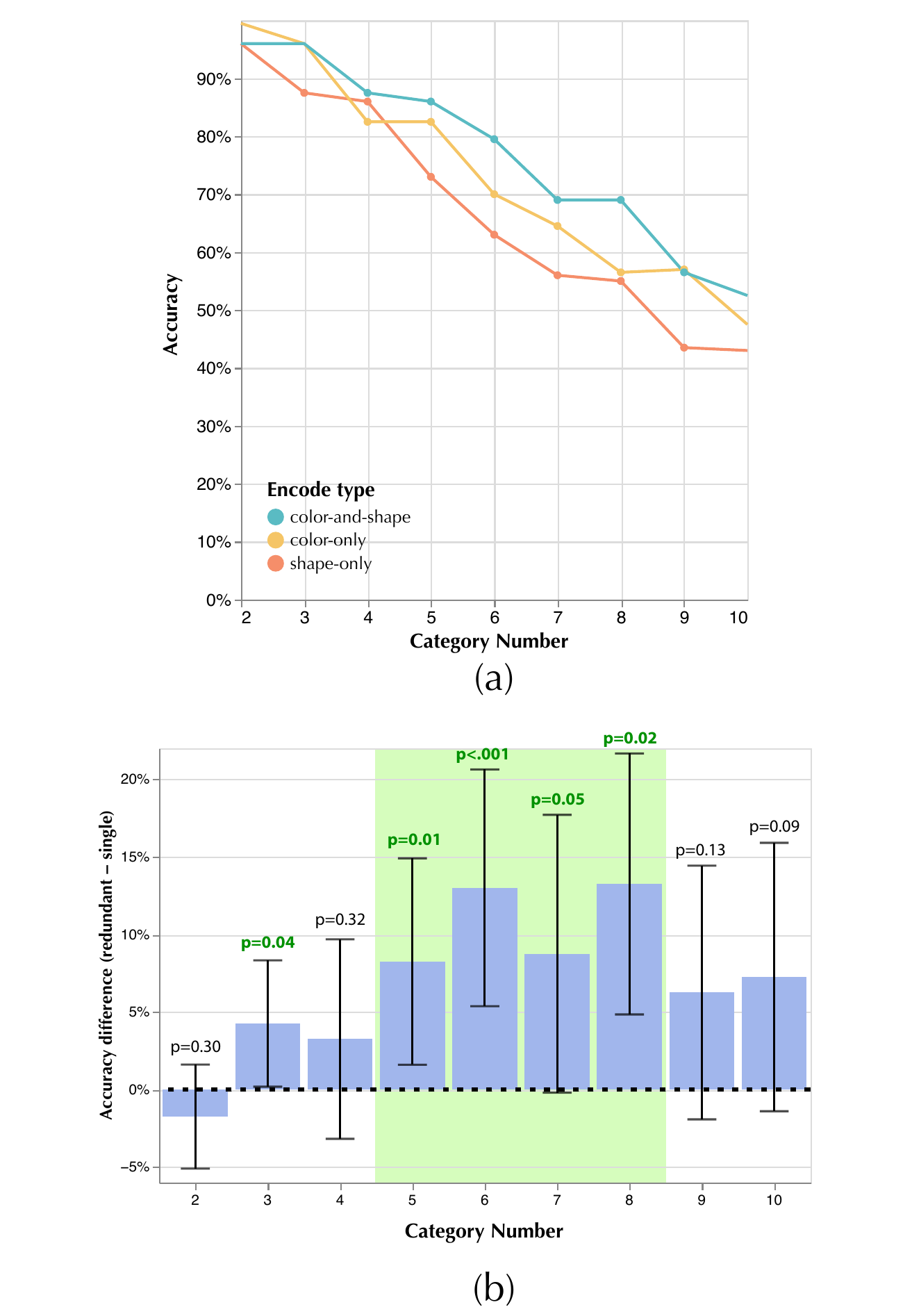} 
\vspace{-0.5em}
\caption{Results from Experiment 1. (a) Line chart showing average accuracy across category sizes (2 to 10), separated by scatterplots using color-only, shape-only, and color-and–shape (redundant) encodings. (b) \add{Bar chart with 95\% CI showing the accuracy difference between redundant and non-redundant encodings
across category numbers. The green background highlights the category range where redundant encoding provides the largest performance gains. For each category number, we conducted a Welch’s t-test; statistically significant differences are shown in bold green text, indicating conditions where redundant encoding significantly outperforms single-channel encodings.}}
\vspace{-0.5em}
\label{fig:result-exp1}
\end{figure}
Our analysis revealed that both category numbers ($F(8, 792) = 116.35, p < .0001$) and encoding types ($F(2, 198) = 27.32, p < .0001$) significantly affected correlation comparisons (\autoref{tab:exp1}). 
\autoref{fig:result-exp1} (a) displays the average accuracy separated by encoding type. Notably, redundant color-and-shape achieved the highest average accuracy at 76.9\% (95\% CI=[74.9, 78.8]), representing an improvement of 4\% over color-only encoding (72.9\%, 95\% CI=[70.8\%, 74.9\%]) and 9.9\% over shape-only encoding (67\%, 95\% CI=[64.8\%, 69.2\%]).

To better understand how redundancy benefits vary across category numbers, we combined color-only and shape-only into a single non-redundant baseline in \autoref{fig:result-exp1} (b). On average, redundant encodings achieved an accuracy of 76.9\% compared to 69.4\% for non-redundant encodings, with the largest difference observed at 8 categories (13.2\%) and the smallest at 2 (-1.75\%). While redundancy provided some benefit for three or more categories, redundant encodings improved performance most consistently when the number of categories ranged from 5 to 8.

\add{To further quantify these differences, we conducted Welch’s \textit{t}-tests at each category number to compare the redundant encodings against non-redundant encodings (Figure \ref{fig:result-exp1} (b)). The results show that redundant encodings significantly outperformed non-redundant encodings for 3 categories ($t=2.06, p=0.04$), 5 categories ($t=2.44, p=0.01$), 6 categories ($t=3.36, p<.001$), 7 categories ($t=1.93, p=0.05$), and 8 categories ($t=3.11, p=0.02$). No significant differences were observed at 2 ($t=-1.03, p=0.30$), 4 ($t=0.99, p=0.32$), 9 ($t=1.51, p=0.13$), or 10 ($t=1.65, p=0.09$) categories. These findings align with the trend in Fig. \ref{fig:result-exp1}(a) that redundancy is most beneficial when visual complexity is moderate (5–8 categories), but, while still beneficial, redundancy may be less helpful when category numbers are too low or too high.} Overall, these results support \textbf{H$_1$}: applying redundant encodings enhances performance on correlation estimation tasks, but this benefit may vary depending on the number of categories.

\section{Experiment Two: Interactions between Color Palettes and Shape Palettes}
\label{sec-exp2}

While Experiment 1 demonstrated that redundant encodings can enhance people's performance in interpreting scatterplots, it remains unclear whether the specific method of creating redundant encodings also impacts task accuracy.  In other words, the way designers pair colors and shapes may influence people’s ability to distinguish between classes. Smart et al.~\cite{smart2019measuring} demonstrated perceptual interactions between colors and shapes, highlighting the importance of their pairing. However, their study focused solely on single shapes and small changes in color rather than categorical designs in multi-class scatterplots, which require reasoning over larger numbers of marks and leveraging greater color differences. In Experiment 2, we investigate how redundant coding palettes influence performance. We use six shape palettes generated from \textit{Shape It Up}~\cite{tseng2024shape} and the four color palettes from Experiment 1 to measure the influence of color-shape pairings. 

We hypothesized that: 
\noindent\textbf{H$_{2}$:} \textbf{
mappings between color palettes and shape palettes impact participants' accuracy in correlation estimation.} Because shapes differ considerably in their colored areas (e.g., filled circles versus unfilled circles), different color attributes may either enhance or hinder performance. For example, more similar colors within a palette may be robust if paired with filled shapes that have more overall color area, but less so with thin, line-based shapes with less color area \cite{smart2019measuring}. If the mapping between color and shape palettes does not matter, we expect that the relative performance of redundant palettes changes as a function of each channel's individual performance
(e.g., color palette X always outperforms color palette Y if X and Y are mapped to the same shape palette). However, if mapping matters, we expect that the choice of both color and shape palette will change their relative effectiveness.

\begin{figure}[htbp] 
\centering
\includegraphics[width=0.35\textwidth]{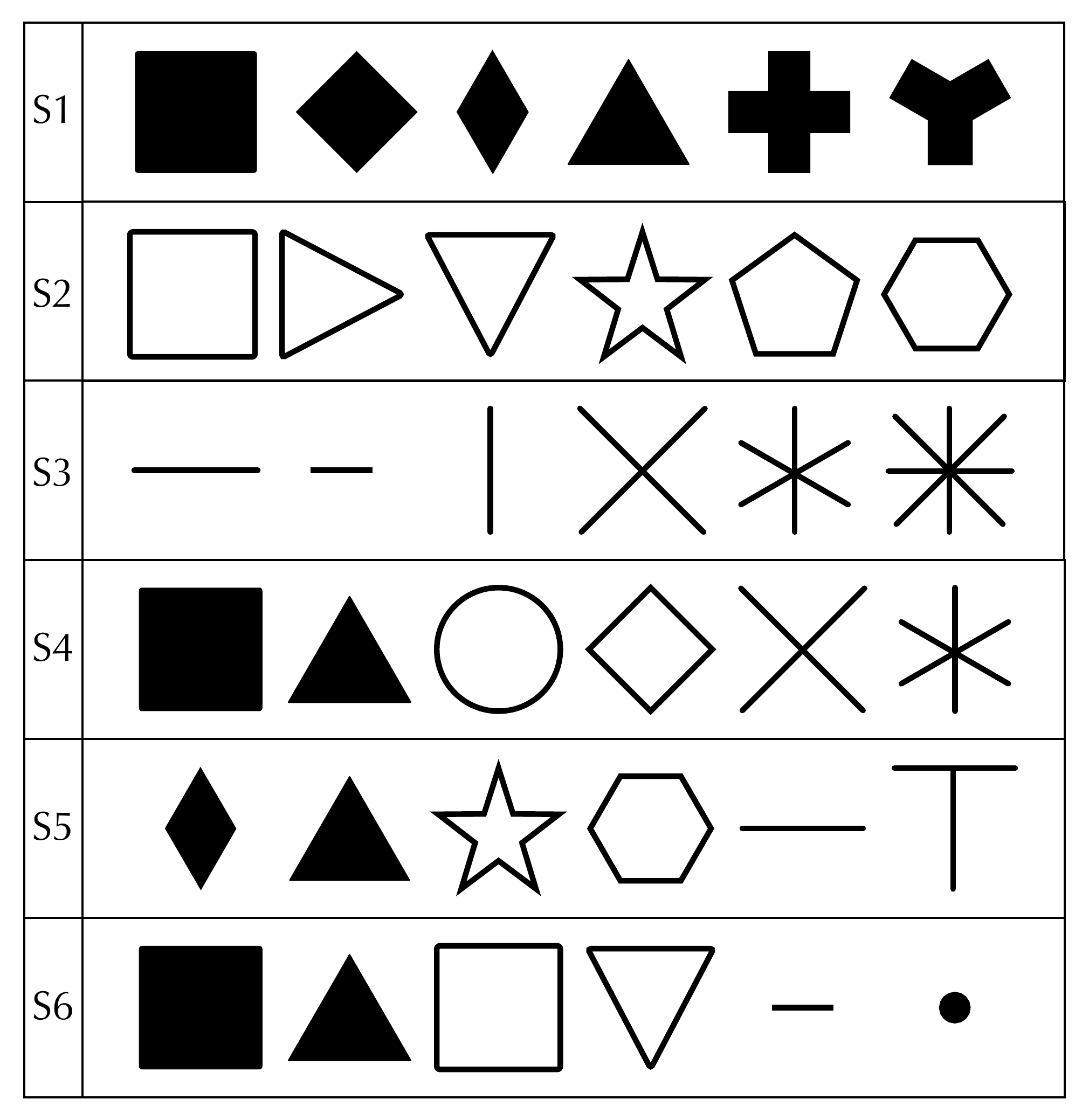} 
\caption{The six shape palettes used in Experiment 2. Each palette consists of six shapes and was generated using Shape It Up~\cite{tseng2024shape}.}
\vspace{-1em}
\label{fig:shape-palettes}
\end{figure}

\subsection{Experiment Design}
\begin{figure}[htbp] 
\centering
\includegraphics[width=0.45\textwidth]{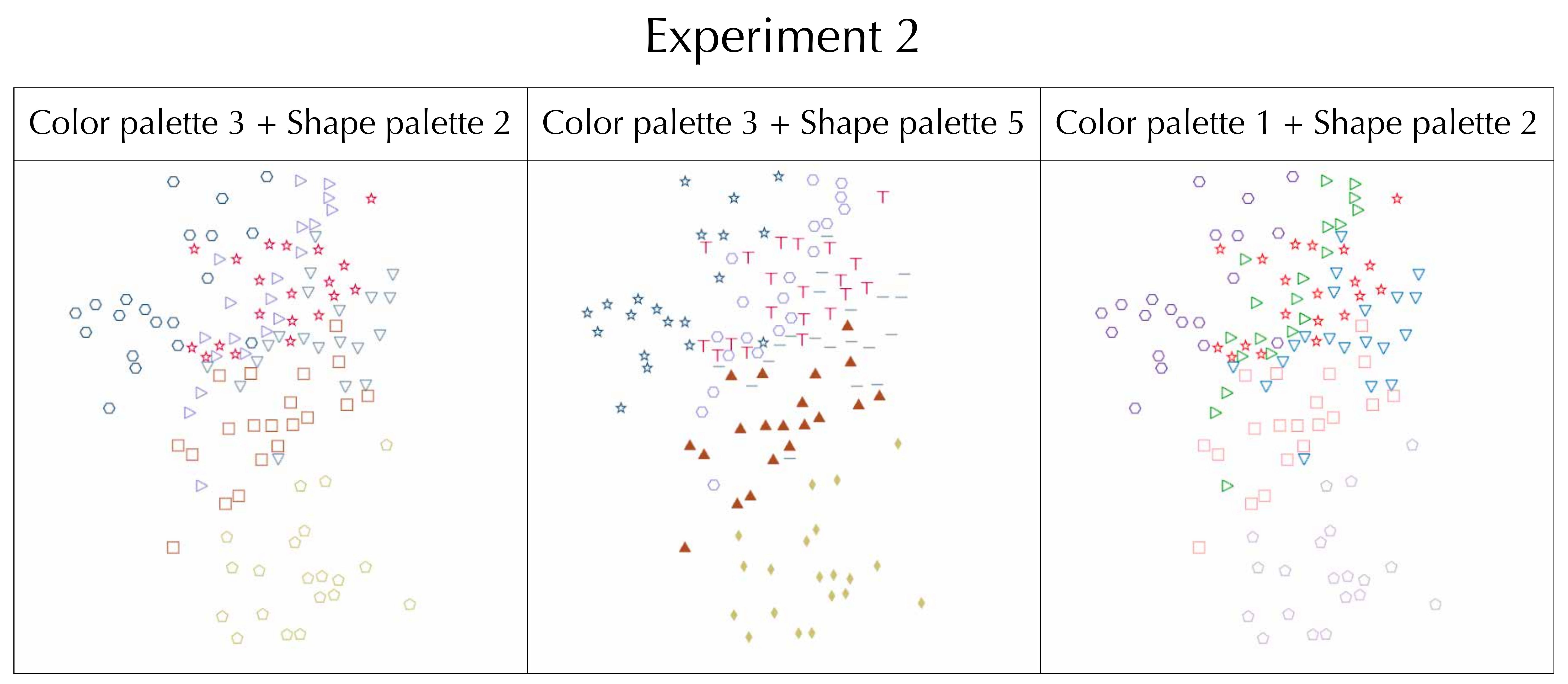} 
\caption{Example stimuli from Experiment 2: three scatterplots using different combinations of color and shape palettes, shown with the same data distribution and 6 categories. Correlation comparison tasks were used in Experiment 2.} 
\vspace{-1em}
\label{fig:stimuli_exp2}
\end{figure}
\subsubsection{Task \& Stimuli Generation}
We used the same task, point distribution, and number of points as in Experiment 1. As shown in \autoref{fig:color-palette} and \autoref{fig:shape-palettes}, we selected 4 color palettes and 6 shape palettes. We elected to study mapping effects at the scale of palettes to reflect design practices used in current visualization authoring tools where people choose palettes which the tool then pairs rather than selecting individual color-shape pairings, which we investigate in more detail in Experiments 3 and 4. The color palettes were chosen for their distinct color attributes (e.g., $L^*$ magnitude), consistent with Experiment 1. The shape palettes were generated using Shape It Up~\cite{tseng2024shape} and included one filled shape \add{palette}, one unfilled shape \add{palette}, one open shape \add{palette}, and the three highest-recommended palettes selected from a pool of 39 shapes. 

We fixed the number of categories at six, since this value is the midpoint of our tested range in Experiment 1 (i.e., between 2 and 10) and exhibited some of the most prominent improvements from redundant encodings. By combining the 6 shape palettes (\autoref{fig:shape-palettes}) with the 4 color palettes (\autoref{fig:color-palette}), we generated 24 combinations of color-and-shape redundant palettes. Within color-and-shape palettes, we randomly paired colors and shapes, with one color assigned to one shape, representing a unique category. Finally, we generated 10 stimuli for each of the 24 color-and-shape palettes, resulting in a total of 240 stimulus designs. \add{These were then evenly divided into five task groups, each containing all 24 palettes with two 
stimuli for each color-and-shape palette combination.}

\subsubsection{Procedure \& Participants}
We followed the same general procedure as in Experiment 1. Each participant completed 51 trials (48 formal trials and 3 engagement checks), presented sequentially in a random order. We recruited 61 participants under the same criteria as in Experiment 1. Six participants who failed the engagement checks were excluded from the analysis, resulting in 
55 participants (24 female, 31 male; age range: 28–65), with 11 participants per task group. On average, each participant spent 10 minutes on the experiment \add{and was compensated \$1.60 for their time}.

\begin{table}[htbp] 
\centering
\vspace{-0.5em}
\caption{ANOVA results for color palette and shape palette. Significant effects are indicated by \textbf{bold} text.}
\vspace{-0.5em}
\includegraphics[width=1\linewidth]{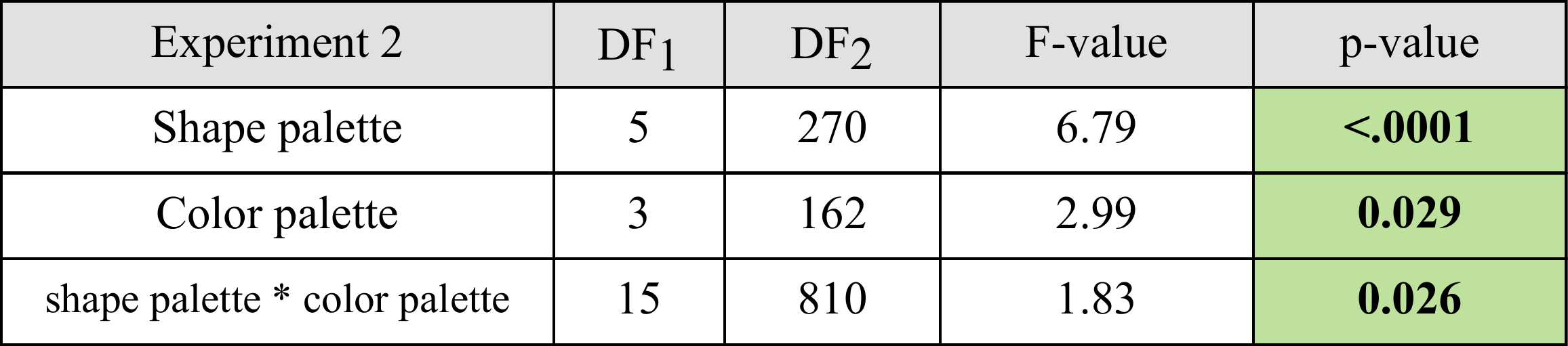} 
\vspace{-0.5em}
\label{tab:exp2}
\end{table}

\begin{figure}[htbp] 
\centering
\vspace{-0.2em}
\includegraphics[width=.48\textwidth]{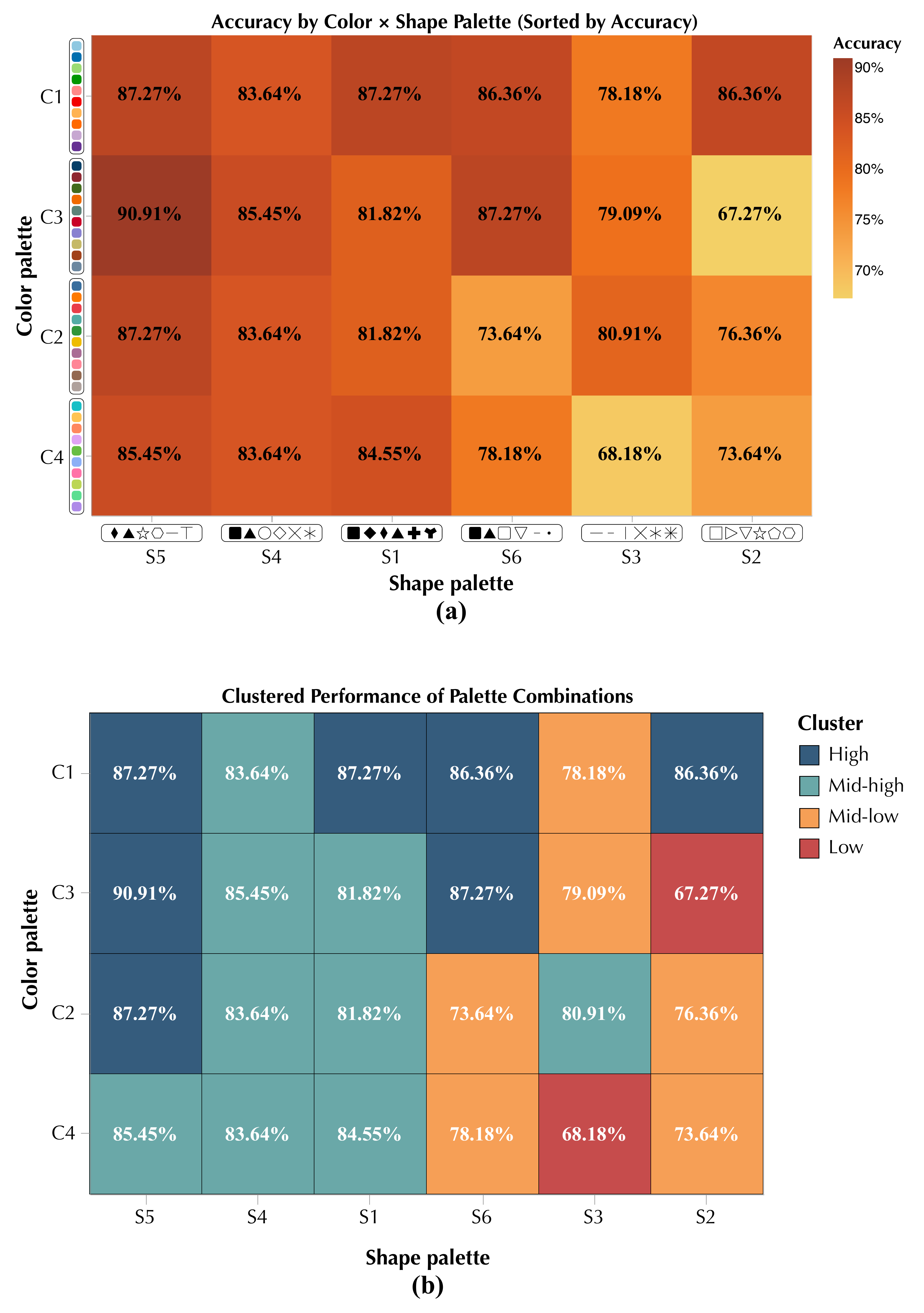} 
\vspace{-1em}
\caption{(a) Heatmap showing the average accuracy for each combination of color and shape palettes. The x-axis lists the six shape palette IDs and the y-axis lists the four color palette IDs (corresponding to \autoref{fig:shape-palettes} and \autoref{fig:color-palette}, respectively). Both axes are sorted by average accuracy to facilitate comparison across palette combinations. Each cell displays the mean task accuracy for the associated palette pairing. As shown, the same color palette can produce different accuracies depending on the shape palette with which it is paired, and vice versa, illustrating strong palette-mapping effects. \add{(b) Clustered visualization of the same color–shape combinations using a four-cluster Ward hierarchical solution, clustering color-and-shape palette combinations based on performance. Each cell is assigned to one of four performance clusters—High, Mid-high, Mid-low, or Low—revealing distinct families of palette pairings. The cluster structure highlights that neither color nor shape palettes are consistently effective on their own: several palettes appear in both high- and low-performing clusters depending on the pairing.}} 
\label{fig:result-exp2}
\end{figure}

\subsection{Results}

We analyzed how the choices of color palettes, shape palettes, and their combinations influence response accuracy using a two-factor ANOVA and Tukey's HSD for posthoc analysis. Our analysis revealed that 
color palettes, shape palettes, and their pairings all significantly affected correlation comparisons (see \autoref{tab:exp2}). \autoref{fig:result-exp2} (a) displays a heatmap where the x-axis represents six different shape palettes and the y-axis represents four different color palettes, with cell values indicating the average task accuracy for each color-shape redundant encoding. The highest average accuracy (90.91\%) was achieved with color palette No. 3 paired with shape palette No. 5, whereas the lowest accuracy (67.27\%) was also observed with color palette No. 3 but when paired with shape palette No. 2. 

The overall accuracy on the correlation comparison task was 81.59\% (95\% CI=[80.11\%, 83.07\%]). The individual average accuracies for the six shape palettes, in the order shown in \autoref{fig:shape-palettes}, were 83.86\%, 75.9\%, 76.59\%, 84.09\%, 87.72\%, and 81.36\% with $\sigma=4.21\%$. For the four color palettes, the individual average accuracies were 84.84\%, 80.6\%, 81.96\%, and 78.93\% with $\sigma=2.16\%$ in \autoref{fig:color-palette} order. The variability in accuracy across shape palettes was slightly greater than color palettes.

We calculated the variance of accuracies across all color palettes using the same shape palettes and 
across shape palettes using the same color palettes. The results show that shape palette No. 2 had the largest variance in accuracies across the four color palettes, and color palette No. 3 (Stata/S2) had the largest variance across the six shape palettes. These findings suggest that certain shape or color palettes may be more sensitive to pairing choices and need closer attention during design. Furthermore, our ANOVA results (see \autoref{tab:exp3}) indicate a significant interaction effect between color and shape palettes, supporting \textbf{H$_{2}$}: how we pair color and shape palettes influences task performance. As shown in \autoref{fig:stimuli_exp2} and \autoref{fig:result-exp2} (a), color palette No. 3 (Stata/S2), when paired with shape palette No. 5, achieved the overall highest accuracy. In contrast, pairing the same color palette with shape palette No. 2 (unfilled shapes) resulted in the overall lowest accuracy. On the other hand, pairing shape palette No. 2 with color palette No. 1 (ColorBrewer/Paired) improved accuracy by nearly 20\% over its pairing with 
Stata/S2.

\add{We additionally 
used Ward's method to cluster the palette combinations based on performance, allowing us to examine broader performance patterns across the 24 color–shape combinations (\autoref{fig:result-exp2} (b)). The method resulted in four performance clusters. The resulting clusters show that neither color nor shape palettes perform consistently well across all pairings; several palettes appear in both high- and low-performing clusters. Only one shape palette and no color palettes appeared entirely within any single cluster. This reinforces the interaction effect between color and shape observed in the ANOVA results and further supports \textbf{H$_{2}$}: performance depends on how color and shape palettes are paired, not just the performance of the palettes in isolation.}

We further examined how color and shape palette attributes influence task performance in redundant pairings. Our findings show that the unfilled (shape palette No. 2) and open (shape palette No. 3) shapes resulted in lower accuracies when paired with color palettes characterized by either high lightness ($L^*$, palette No. 4--Carto/Pastel) or low lightness (palette No. 3--Stata/S2). In contrast, these shape palettes achieved higher accuracies when paired with color palettes exhibiting greater perceptual distances (palettes No. 1--ColorBrewer/Paired and 2--Tableau). Overall, mixed-shape-type (filled+unfilled+open) palettes (palettes No. 4, 5, and 6) achieved higher average accuracies, particularly in improving accuracies with lighter or darker color palettes (palettes No. 3--Stata/S2 and 4--Carto/Pastel).

\section{Experiment Three: Pairwise Distance Matrix for Colors}
\label{sec-exp3}

Experiments 1 and 2 found that redundant encodings not only enhance task performance but that the specific 
color–shape palette pairings also influence accuracy. %
To make these findings actionable, we aimed to develop an empirically-grounded model that generates effective redundant color-and-shape palettes based on the number of categories visualized.
To achieve this goal, we conducted two separate experiments: one to build a pairwise accuracy matrix for colors (Experiment 3) to complement existing shape palette design measures and another to explore efficient color-shape pairings (Experiment 4). We separated these experiments because constructing a pairwise matrix for 39 colors and 39 shapes would involve over two million comparisons. Instead, this approach allows us to first identify effective color sets, which can be sampled for the next experiment. While this approach does potentially exclude useful shape-color combinations, it allows our sampling process to focus on empirically-validated color combinations, increasing the likelihood that the modeled space captures high-performing palettes. 

While past work has compared the effectiveness of different categorical palettes \cite{tseng2023evaluating,tseng2024revisiting,gramazio2016colorgorical,lu2020palettailor}, these studies focus on comparing entire palettes rather than specific pairwise color relations. Our objective is to collect data to model how specific color pairings influence categorical analysis. 
We hypothesized that: 
\textbf{H$_{3}$:} \textbf{The relationships between colors will influence people's abilities to compare correlations.} Given that various studies~\cite{gramazio2016colorgorical, tseng2023evaluating, tseng2024revisiting} have emphasized the importance of selecting appropriate colors for encoding categorical data in scatterplots, we expect that different color combinations will impact performance. Furthermore, different color combinations may be differently robust to varying numbers of categories~\cite{tseng2023evaluating}.

\begin{figure}[htbp] 
\centering
\includegraphics[width=0.45\textwidth]{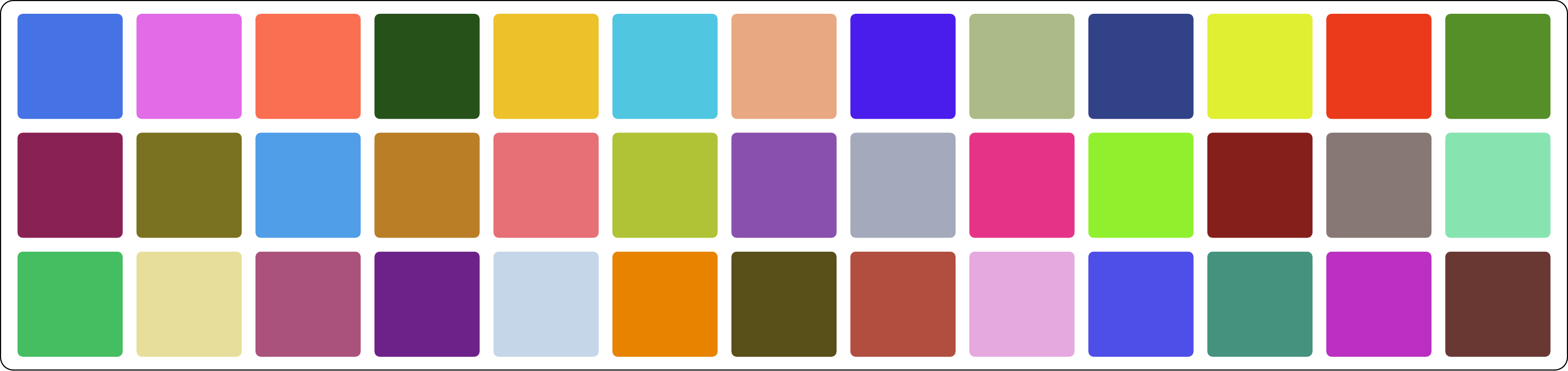} 
\caption{The 39 colors used in Experiment 3. These colors were generated using k-means clustering in the CIELAB color space. Each pair exceeds the just-noticeable difference (JND) value for the mark sizes used in our study.} 
\vspace{-0.5em}
\label{fig:kmeans-palettes}
\end{figure}

\subsection{Experiment Design}
\begin{figure}[htbp] 
\centering
\includegraphics[width=0.45\textwidth]{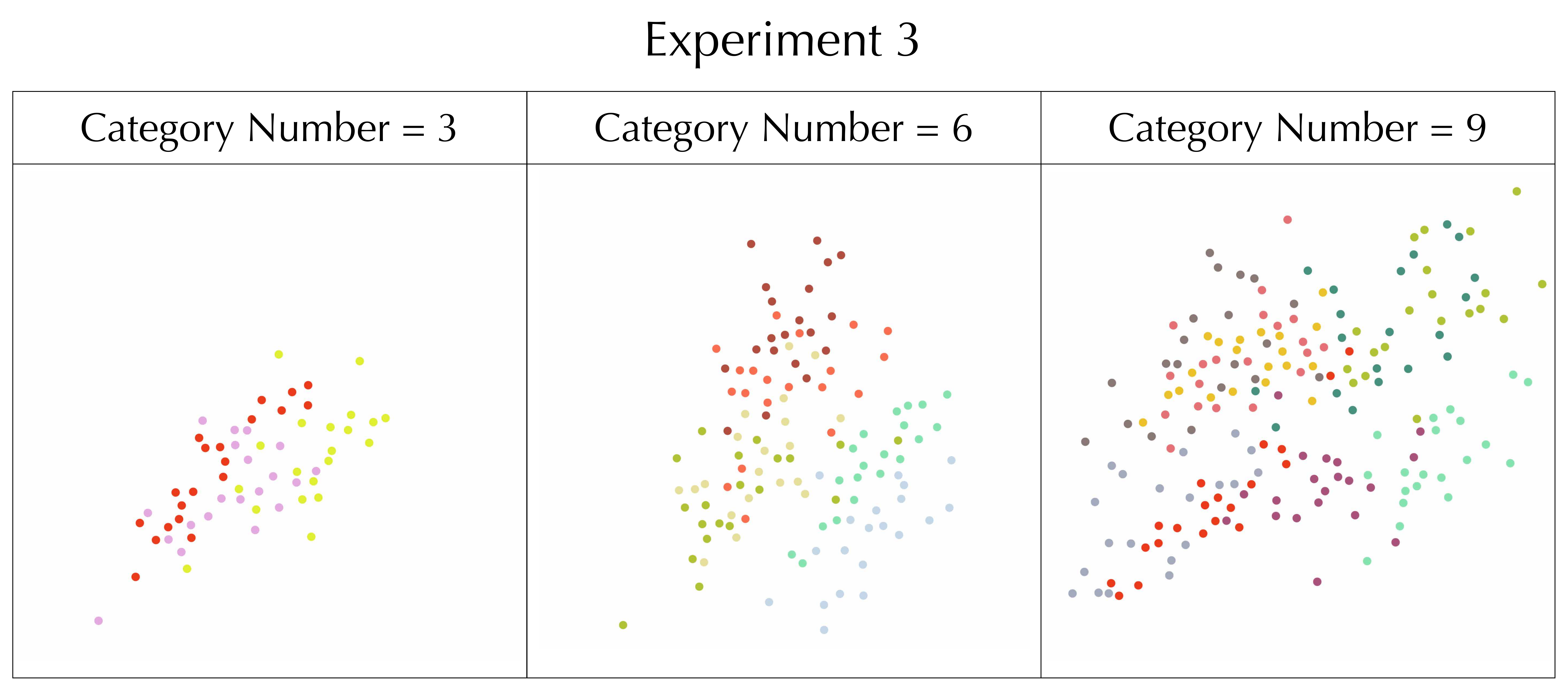} 
\caption{Example stimuli from Experiment 3, showing scatterplots with 3, 6, and 9 categories encoded using color only. This experiment was designed to construct pairwise accuracy matrices for 39 colors, capturing how effectively different color pairs support correlation estimation.}
\label{fig:stimuli_exp3}
\end{figure}
\subsubsection{Task \& Stimuli Generation}
The task, point distribution, and point number were generated using the same methodology as in Experiments 1 and 2. Following the approach in Tseng et al.~\cite{tseng2024shape}, we focused on performance differences between color pairs by constructing a distance matrix. Unlike shapes in a discrete space, colors exist in a continuous space.
Past work has attempted to bin colors into a smaller and more manageable set of readily distinguishable, representative values. However, the suggested colors largely vary---from sets of eight~\cite{wong2010points} to thirteen~\cite{berlin1991basic} to twenty-two~\cite{green2010colour}.
Further, these characteristic colors may fail 
to meet users' needs and preferences for color aesthetics~\cite{schloss2010aesthetics, schloss2013object}, semantics~\cite{schloss2018color}, or other design factors \cite{bartram2017affective}.
Therefore, following prior work~\cite{gramazio2016colorgorical, hong2024cieran}, we 
derive our core colors by mathematically sampling across colorspace to ensure the diversity and representativeness of colors.
To ensure sufficient and evenly distributed colors for our model, we employed k-means clustering across the CIELAB colorspace, as implemented by iWantHue~\cite{iwanthue}. CIELAB is designed to be perceptually uniform, encompasses all perceivable colors, and has been widely used in past visualization research~\cite{gramazio2016colorgorical, szafir2014adapting, smart2019color, ware2018measuring, chen2014visual, wang2008color}.

We first sampled colors within the ranges: $L^* = [25, 100]$, $a^* = [-128, 127]$, and $b^* = [-128, 127]$, using 5 steps for $L^*$ and 2 steps for both $a^*$ and $b^*$. Colors darker than $L^* = 25$ were excluded, as the CIELAB space tends to contain more colors at lower lightness values, which can lead to an unbalanced sample, and design heuristics discourage using colors in this range for visualization \cite{stone2006choosing}. 
This process yielded 38,734 valid color samples. We then applied k-means clustering to extract 200 representative colors. To ensure perceptual distinguishability between color pairs, we used d3-jnd~\cite{gramazio_d3jnd_github, stone2014engineering} to validate just-noticeable differences (JND) between each pair for our target mark size. For each of the 200 colors, we computed the number of valid JND pairings and selected the maximum subset in which all pairs were noticeable differently~\cite{gramazio2016colorgorical}. This step eliminated 163 colors, resulting in a refined set of 37 colors.

To further balance the set, we analyzed color name distribution using C3~\cite{heer2012color}. While the overall distribution was balanced, we observed a lack of orange hues. To address this, we manually added two orange colors from C3~\cite{heer2012color}, bringing the final color set to 39. We also used d3-jnd~\cite{gramazio_d3jnd_github, stone2014engineering} to ensure each color is sufficiently perceptible against our tested white background. The resulting palette is shown in \autoref{fig:kmeans-palettes}, \add{and example stimuli are shown in \autoref{fig:stimuli_exp3}(a).}

We generated the tested color combinations based on the approach proposed by Tseng et al.~\cite{tseng2024shape} for shapes, which ensures a balanced distribution of pairwise samples. Using this method, we created 810 color combination sets (90 color combinations $\times$ 9 category counts). These sets were then organized into 15 task groups, with each group containing an equal distribution across category counts (6 color combination sets $\times$ 9 category counts = 54 stimuli per group).

\subsubsection{Procedure \& Participants}
The recruitment criteria and general procedure were consistent with those used in Experiments 1 and 2  (see Section \ref{sec-exp1}). We recruited 167 participants through MTurk, excluding 17 who failed the engagement checks. The final analysis included data from 150 participants (107 male, 43 female; ages 21 to 65), with 10 participants assigned to each task group.  The study took approximately 10 minutes to complete \add{and each participant was compensated \$1.60}. 

\subsubsection{Analysis}
Our primary dependent measure was pairwise accuracy, a metric employed in previous work~\cite{gramazio2016colorgorical, tseng2024shape}. This metric was computed by evaluating the accuracy of each scatterplot containing a given pair of colors. For example, if a scatterplot having blue, orange, and green was answered correctly, the pairs blue–orange, blue–green, and orange–green each received one correct count, \add{whereas an incorrect response contributed zero}. Pairwise accuracy for each color pair was then calculated as the number of correct responses divided by the total number of trials in which that pair appeared. 
\add{Because the correlation estimation task requires participants to visually identify all categories and compare their correlations simultaneously, every category in the stimulus contributes to the overall judgment. Misattributing marks to the wrong category or failing to find a mark of a given category can change the perceived correlation \cite{barras2017target}. For example, failing to find a cluster of outlying blues because it becomes masked by a cluster of green points will increase the perceived correlation of blue and may, as a result, change the relative perceived correlation of all three categories. This scoring approach therefore captures how well each color-shape pair supports category differentiation within a multi-class context, while avoiding the need for an impractically large number of isolated pairwise trials.}
We conducted a three-factor ANOVA to examine the effects of category number, color pair, and color attributes on accuracy. A summary of the results is presented in \autoref{tab:exp3}.

\begin{table}[htbp] 
\centering
\vspace{-0.5em}
\caption{ANOVA results for category size and color pair. Significant effects are indicated by \textbf{bold} text.}
\vspace{-0.5em}
\includegraphics[width=1\linewidth]{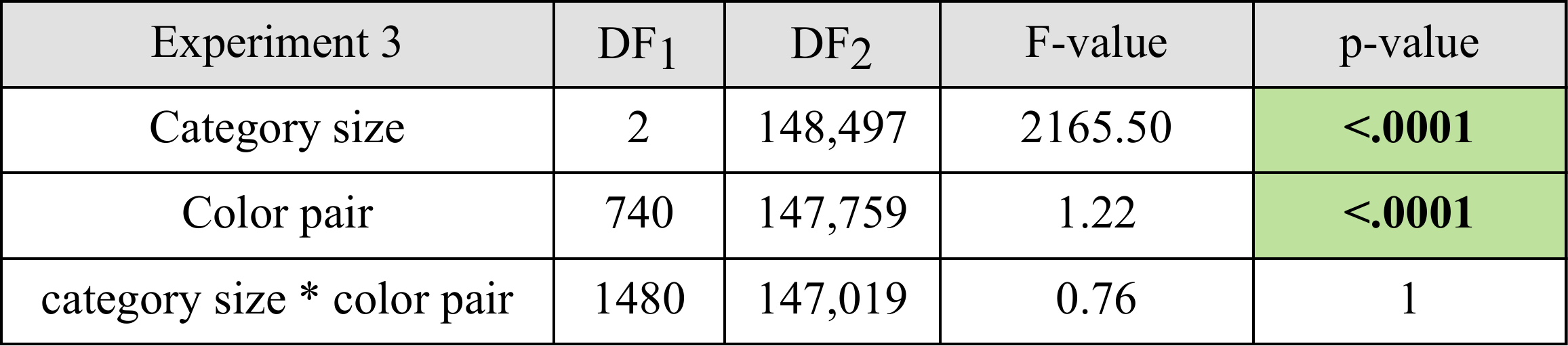} 
\vspace{-2em}
\label{tab:exp3}
\end{table}

\begin{table}[htbp] 
\centering
\caption{ANOVA results for color attributes. Significant effects are indicated by \textbf{bold} text.}
\vspace{-0.5em}
\includegraphics[width=1\linewidth]{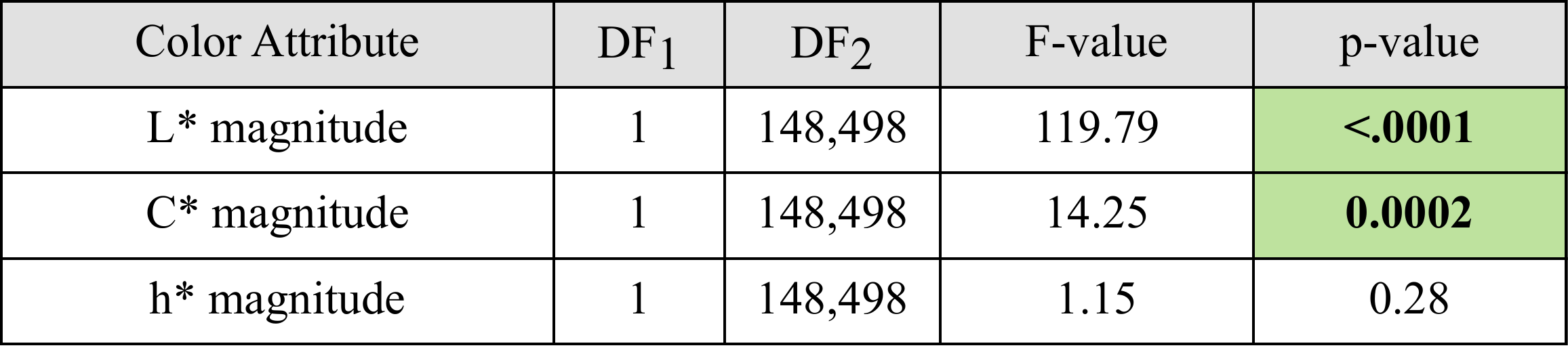} 
\label{tab:exp3-2}
\end{table}

\subsection{Results}
As shown in \autoref{tab:exp3}, our analysis revealed significant effects of both category size ($F(2, 148497) = 2165.5, p < .0001$) and specific color pairs ($F(740, 147759) = 1.22, p < .0001$) on performance in the correlation comparison task. The overall accuracy was 78.14\%. Accuracy varied across category sizes, with small ($N=2, 3, 4$), medium ($N=5, 6, 7$), and large ($N=8, 9, 10$) categories having accuracies of 91.59\%, 78.33\%, and 64.48\%, respectively, reflecting a decrease in accuracy as the number of categories increased. We summarized these results using heatmaps presented on an interactive \href{https://catpaw-categorical-palette.web.app/pair.html?group=all}{website\footnote{\add{\url{https://catpaw-categorical-palette.web.app/pair.html?group=all}}}}, allowing detailed exploration of accuracy differences across color pairs.
We failed to find a significant effect of \textit{Perceptual Distance};
however, this is likely due to the fact that we already optimized distance during the stimulus generation by ensuring color differences exceeded necessary JNDs. 
Above a certain threshold, the differences between colors are large enough 
that making them larger does not meaningfully improve discriminability \cite{heer2012color}.

Overall, the pairwise accuracy ranged from 56.84\% to 80.5\%, with a standard deviation of 3.56\%. Our results support \textbf{H$_{3}$}: color pairs impact task performance. \add{Examples of best-performing color palettes are shown in the \href{https://osf.io/4up7j/?view_only=43e5681b71fc4fdb92617b8cef27c6c0}{OSF Supplements}}. We further examined the relationship between various color attributes and pairwise accuracy 
using CIELCh~\cite{zeileis2009escaping}, which is 
commonly used for color palette design~\cite{stone2006choosing, gramazio2016colorgorical} and comprises a polar representation of the CIELAB colorspace. For all 741 color pairs, we computed the $L^*$, $C^*$, and $h^*$ (Lightness, Chroma, Hue) magnitude/difference,
as well as the name difference~\cite{heer2012color}, composite difference (perceptual distance~\cite{sharma2005ciede2000}) and decomposed ($L^*$, $a^*$, $b^*$) from CIELAB.
\autoref{tab:exp3-2} presents the exploratory ANOVA results. 
The magnitudes in $L^*$ and $C^*$ significantly affect task performance, whereas perceptual distance; $L^*$, $a^*$, $b^*$, and $h^*$ differences; and name differences appear to be less influential (shown in \href{https://osf.io/4up7j/?view_only=43e5681b71fc4fdb92617b8cef27c6c0}{OSF Supplements}), likely due to all selected colors being sufficiently perceptually distinct by design. These findings are consistent with previous work~\cite{tseng2023evaluating}, suggesting that lightness and chroma magnitudes are 
critical 
for categorical perception.

\section{Experiment Four: Redundant Encoding Matrices}
\label{sec-exp4}
Experiment 2 revealed that the pairing of color and shape palettes significantly affects task performance. Therefore, rather than simply combining the best-performing color and shape palettes, effective redundant encodings must be designed with redundancy in mind. We designed Experiment 4 to explicitly measure the effectiveness of different color–shape pairings.
We hypothesized the following:
\noindent\textbf{H$_{4}$: (a) How colors and shapes are paired will influence correlation comparison accuracy, and (b) combining the best-performing color pairs and shape pairs does not necessarily produce optimal encodings.} As Experiment 2 highlighted the importance of pairing at the palette level, we extend this assumption to the individual marker level (e.g., blue triangle vs. blue square). Consequently, although we selected high-performing palettes, we do not expect that random combinations of color and shape elements from those palettes will automatically result in optimal redundant encodings.
 
\subsection{Experiment Design}
\begin{figure}[htbp] 
\centering
\includegraphics[width=0.45\textwidth]{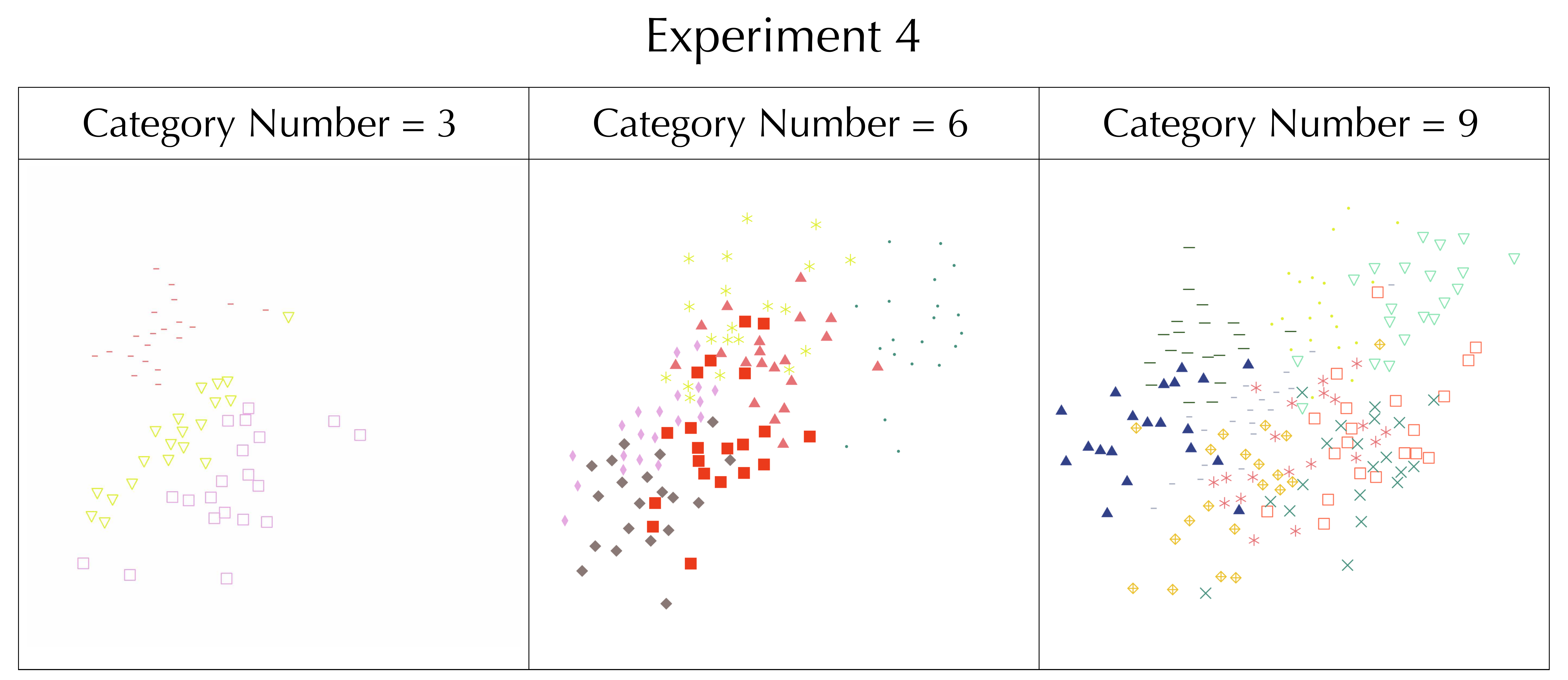} 
\caption{\add{Example stimuli from Experiment 4, showing scatterplots with 3, 6, and 9 categories encoded using redundant color–shape combinations, where each category is represented by a unique pairing of a color and a geometric shape. This experiment was used to construct pairwise accuracy matrices for all 39 colors $\times$ 39 shapes to assess how combined encodings influence perceptual discrimination.}}
\vspace{-1em}
\label{fig:stimuli_exp4}
\end{figure}
\subsubsection{Task \& Stimuli Generation}
We adopted the same task design, point distribution, and point count as in the previous three experiments. Building on the pairwise accuracy matrices obtained in Experiment 3, we developed a palette selection model that takes category number ($N$) as input and outputs high-performing color sets from the pool of 39 colors from Experiment 3. The model randomly selects half of the color pool, generates all possible combinations for the given $N$, computes the sum of pairwise accuracies for each combination, and identifies the combinations with the highest average pairwise accuracy. This process is repeated ten times to ensure stability. For each category number ($N=2$–$10$), we selected the top three color palettes using this approach, and paired them with the top three shape palettes generated using ShapeItUp~\cite{tseng2024shape}, resulting in 9 color–shape palette combinations per $N$. 

To expand the range of tested color–shape pairings, we designed a pairing strategy rather than relying on random assignment. Exhaustively testing all permutations of color–shape matches is computationally infeasible for higher category counts (e.g., $N=5$ yields $5! = 120$ possible permutations). For $N=2$ and $N=3$, we included all 2 and 6 permutations, respectively. For $N=4$–$10$, we selected 13 representative permutations for each color–shape palette pair using a greedy algorithm which selects one palette with the lowest cosine similarity compared to selected ones at a time, ensuring a diverse sampling of redundant encodings. This results in 99 (2+6+7$\times$13) redundant palettes $\times$ 9 (3 color palettes $\times$ 3 shape palettes) palette combinations, resulting in $891$ stimulus designs. We divided these 891 stimuli into 15 task groups, each containing 60 stimuli with an approximately uniform distribution of category number and source palette. \add{Example stimuli are shown in \autoref{fig:stimuli_exp4}(b).}

\subsubsection{Procedure \& Participants}
The recruitment criteria and overall procedure matched those used in Experiments 1, 2, and 3 (see Section \ref{sec-exp1}). A total of 170 participants were recruited via MTurk, with 20 excluded for not passing the engagement checks. This left 150 participants (115 male, 35 female; aged between 28 and 65) whose data were included in the final analysis, with 10 participants assigned to each task group. On average, the study took 10 minutes to complete, \add{and participants were compensated \$2.90 for their time.} 

\subsubsection{Analysis}
We used task accuracy as the primary dependent variable and conducted two main analyses: (1) individual color–shape pairing accuracy (e.g., yellow triangle, blue cross) and (2) pairwise marker accuracy (e.g., red circle + green square). For the individual color–shape pairing analysis, we calculated accuracy by evaluating tasks in which a specific color–shape marker appeared. For instance, if a scatterplot containing red circles, green squares, and orange diamonds was answered correctly, each of those three markers received one correct count. Accuracy for each color–shape pairing was then computed as the ratio of correct responses to the total number of appearances. 

For pairwise accuracy, we adopted a similar approach but focused on specific pairs of color–shape markers. Accuracy was calculated based on the frequency of correct responses when a particular pair appeared together in a task. Both analyses were stratified by category size, grouped into three bins based on the number of categories---small ($N=2-4$), medium ($N=5-7$), and large ($N=8-10$)---to 
enable sufficient sample sizes across conditions. We conducted an ANOVA to examine the effect of individual color–shape pairings on correlation comparison task performance. Additionally, to investigate whether simply combining top-performing color pairs and top-performing shape pairs leads to effective redundant encodings, we performed a secondary ANOVA %
to compare pairwise accuracies of both colors and shapes against their joint combinations.

\begin{figure}[htbp] 
\centering
\includegraphics[width=0.45\textwidth]{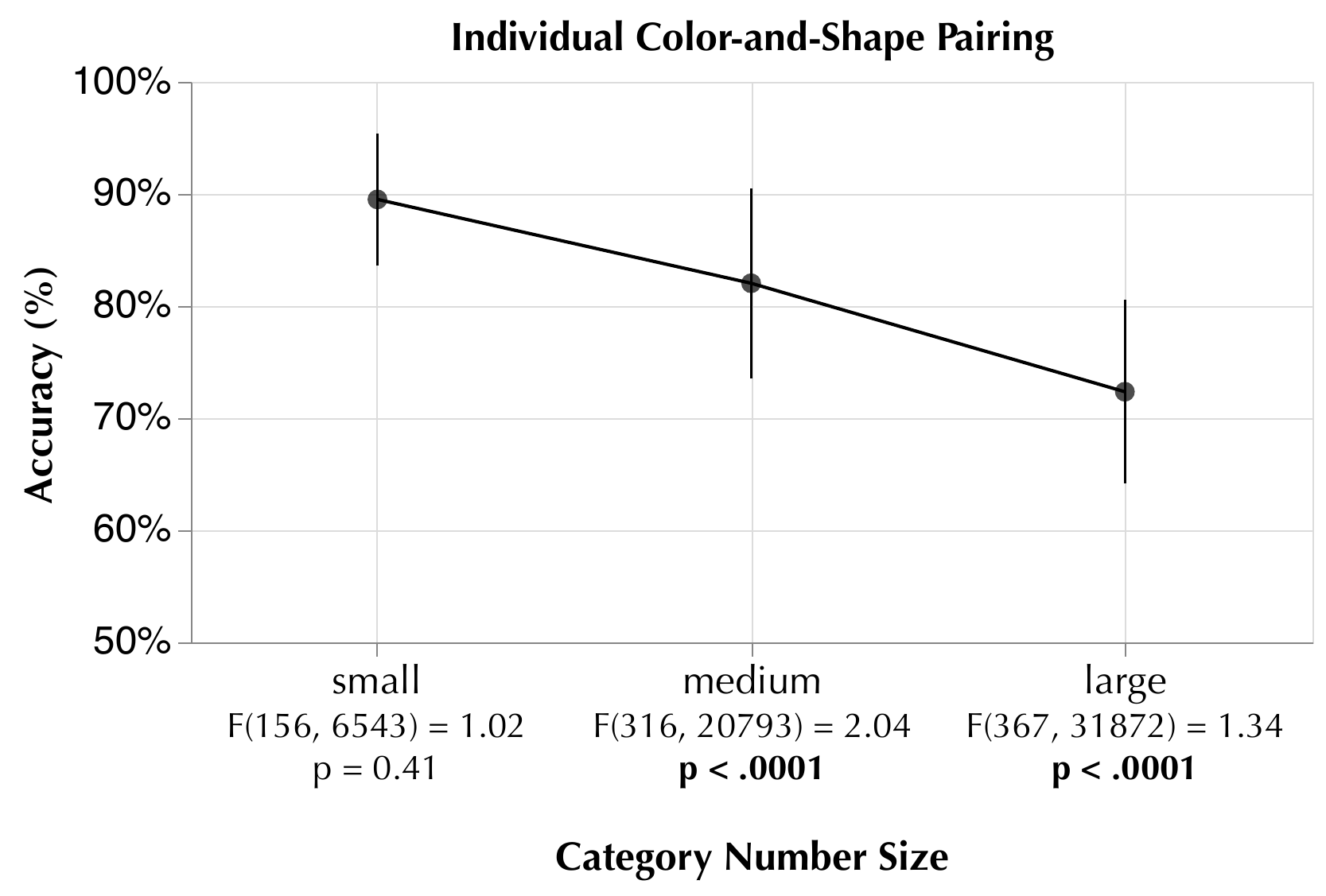} 
\caption{Results from Experiment 4. Average accuracy for individual color–shape pairings across three category size groups (small, medium, and large). Error bars represent standard deviation. ANOVA results indicate significant effects of individual color–shape pairings on task accuracy for medium and large category sizes. These indicate how color-shape paired will impact the task accuracy, particularly with category numbers larger than four.}
\vspace{-0.5em}
\label{fig:result-exp4}
\end{figure}

\subsection{Results}
As shown in \autoref{fig:result-exp4}, average accuracy decreases as category size grows, with an overall accuracy of 79.9\%. Our analysis revealed that the way in which colors and shapes are paired to form a single marker has a significant effect on task performance for medium ($F(316, 20793) = 2.04$, $p < .0001$) and large ($F(367, 31872) = 1.34$, $p < .0001$) category sizes. However, no significant effect was found for small category sizes ($N = 2-4$), ($F(156, 6543) = 1.02$, $p = 0.41$). \autoref{fig:result-exp4} also presents the standard deviations of individual color–shape pairings across category sizes, illustrating greater variance in accuracy for medium and large categories compared to smaller ones. These findings partially support \textbf{H$_{4}$} (a): how colors and shapes are paired influences the accuracy, though this effect appears less prominent for small category sizes. \add{Examples of best-performing color-and-shape palettes are shown in the \href{https://osf.io/4up7j/?view_only=43e5681b71fc4fdb92617b8cef27c6c0}{OSF Supplements}}.

To further investigate the importance of mapping choices, we examined the relationship between pairwise marker accuracy in this experiment and the color/shape pairwise accuracies derived from previous models. 
For each marker pair (e.g., red circle and green square), we computed its task accuracy and retrieved corresponding color pair (e.g., red + green) and shape pair (e.g., circle + square) accuracies from previous matrices. ANOVA results indicate that combining top-performing color pairs and shape pairs simultaneously ($F(1, 4221) = 36, p < .0001$) produce effective redundant encodings only within medium category sizes ($N = 5-7$), where redundant encodings are most effective. Across all category sizes, 
pairwise shape accuracy only showed a significant effect for large category numbers ($F(1, 10053) = 4.18, p = 0.04$), and we did not find any significant effects of pairwise color accuracy. These findings partially support \textbf{H$_{4}$} (b): combining best-performing color and shape pairs does not always lead to optimal encodings, though such combinations appear to be more effective for 5 to 7 categories.

\section{CatPAW: Categorical Palette Automation Wizard}
\label{sec-tool}
\begin{figure*}[htbp] 
\centering
\vspace{-0.2em}
\includegraphics[width=.8\textwidth]{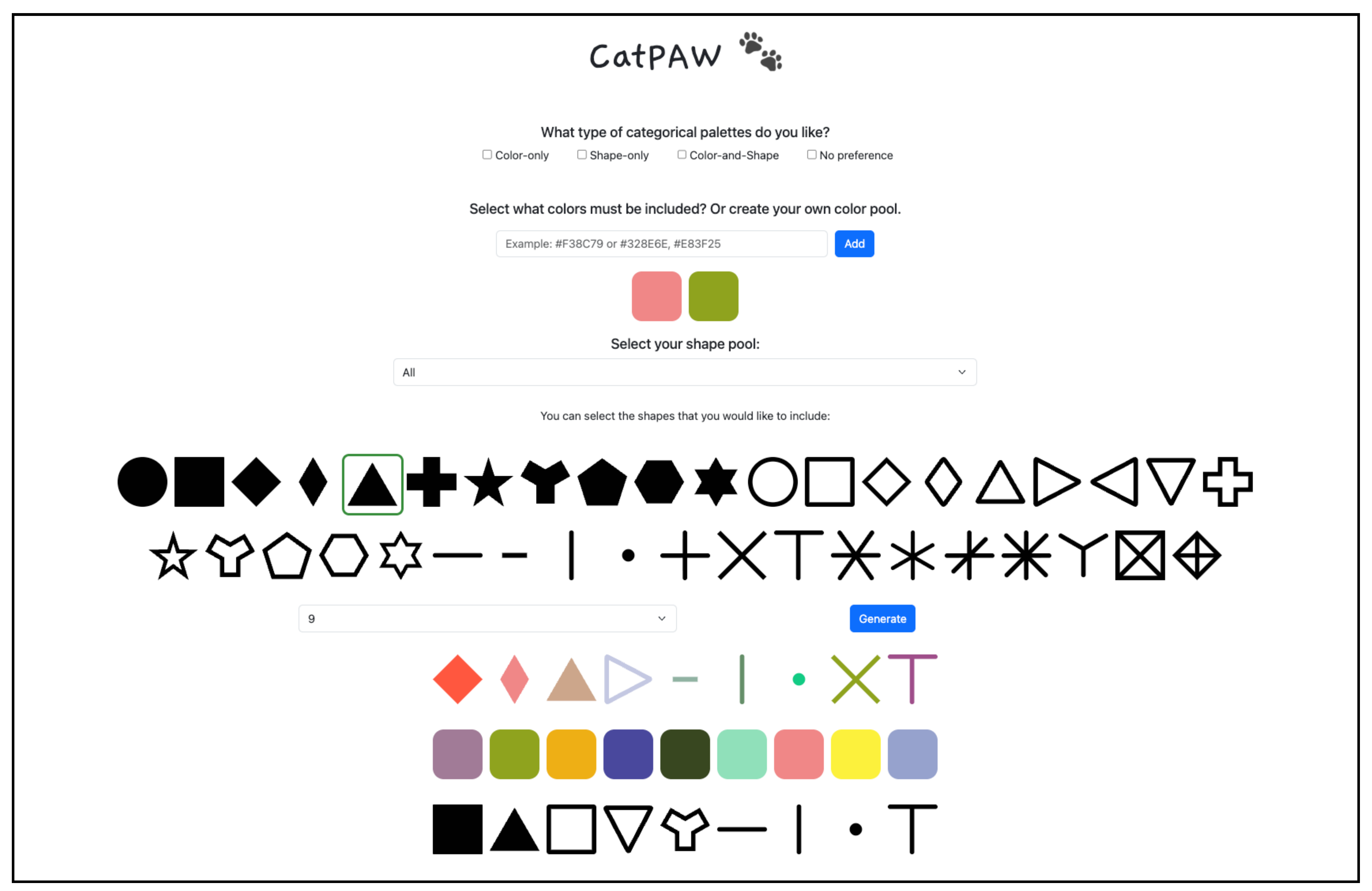} 
\vspace{0em}
\caption{The \emph{CatPAW} interface. Users can specify their preferred palette type (color-only, shape-only, color–and-shape, or no preference), select required colors and/or shapes, and enter the desired number of categories. \emph{CatPAW} then generates candidate palettes, which users can iteratively refine by swapping out unwanted elements with the next highest ranked option.}
\label{fig:catPAW}
\end{figure*}

\subsection{Comprehensive Categorical Encoding Model}
To translate our experimental findings into an actionable framework, we developed a statistical model to evaluate the effectiveness of categorical encodings using shape and color. This model is grounded in a rich set of empirical data, including four pairwise accuracy matrices for 39 colors derived from 14,850 trials in Experiment 3, 839 individual color–shape pairings, and 14,935 pairwise color–shape accuracy samples from Experiment 4. Additionally, we incorporated pairwise accuracy matrices for 39 shapes from Tseng et al.~\cite{tseng2024shape}
and key findings from Experiment 1 (highlighting the effectiveness of redundant encoding within a moderate category range) and Experiment 2 (showing that mixed shape types generally perform better). 

Given a specified category number and encoding type (color-only, shape-only, or color–and-shape), the model operates in the following way: For color-only or shape-only palettes, the model first samples a manageable subset of candidate palettes from the corresponding set of 39 shapes or colors and computes their overall pairwise accuracy based on the selected category number. It then returns the top-performing palette(s). For redundant (color–and-shape) encodings, the model randomly samples combinations of color and shape palettes from each of the 39 color/shape pools and generates a tractable number of color–shape pairing combinations. Each pairing set is evaluated using individual color-shape pairs and pairwise-level accuracy results from Experiment 4. The top-scoring combinations are further assessed based on their color-only and shape-only pairwise accuracy scores, variance in lightness, and variance in shape types. The final output is a ranked list of color–shape paired palettes with corresponding performance scores.

\subsection{Design Tool for Categorical Palettes}
We implemented our model in a web-based categorical palette recommendation tool (see {\href{https://catpaw-categorical-palette.web.app/}{\textit{CatPAW},}}\footnote{\url{https://catpaw-categorical-palette.web.app/}} shown in \autoref{fig:catPAW}). 
While most visualization tools provide default color or shape palettes for encoding categorical data, they often lack guidance on when to use color-only, shape-only, or redundant (color-and-shape) encodings. Moreover, users may wish to incorporate personal design preferences, such as specific colors or shapes, without sacrificing perceptual effectiveness.
Our findings emphasize that choosing appropriate encodings depends on the number of categories and the interaction between visual channels. To support designers in reasoning across these constraints, \emph{CatPAW} offers a comprehensive palette suggestion framework that helps designers generate effective, customizable palettes backed by empirical data.

\emph{CatPAW} allows users to specify the type of categorical palettes they prefer (color, shape, or redundant), and 
any desired colors, shapes, or color-shape pairs.  Based on the selected category number and user constraints, the tool recommends optimal palette configurations. 
We leverage the fact that color can be expressed in a continuous space to extend the generated palettes to a wider range of colors. 
We expect that colors within a given distance of a tested color will have comparable performance. Therefore, rather than directly selecting subsets from the original 39 colors, the tool computes palettes using these colors and then randomly samples nearby colors that are within a certain perceptual distance ($\Delta E \leq 15$) to the original color to increase design flexibility. Specifically, in CIELAB space we jitter each reference color by sampling lightness within $\pm 5$ ($\Delta L \in [-5,5]$) and chromatic components within $\pm 10$ ($\Delta a, \Delta b \in [-10,10]$). \add{We note that this approach provides an approximation of likely performance across the colorspace; however, future studies should derive a more granular model through additional data collection studies.} Similarly, when users input custom colors, the tool maps them to the closest representative color and proceeds with the palette generation accordingly. \emph{CatPAW} provides three core features:

\vspace{3pt}
\noindent\textit{1.} \textit{Support for three categorical palette types.} Users can select from color-only, shape-only, or color-and-shape categorical palettes or allow the tool to recommend the most effective encoding strategy based on the provided constraints.

\noindent\textit{2.} \textit{Data-driven palette suggestions based on category number.} The underlying model draws on empirical results from 455 participants across four experiments. It generates optimized palettes based on category number and adapts to the user’s design preferences.

\noindent\textit{3.} \textit{Customizable palettes with user-defined constraints.} Users can specify pools of colors or shapes or define required elements that must appear in the output. After palette generation, users can also remove unwanted elements, prompting the tool to substitute them with the next best-performing alternatives.

\subsection{\add{Cross-Measure Validation using the Categorical Encoding Model}}

\begin{figure}[htbp] 
\centering
\vspace{-0.2em}
\includegraphics[width=.45\textwidth]{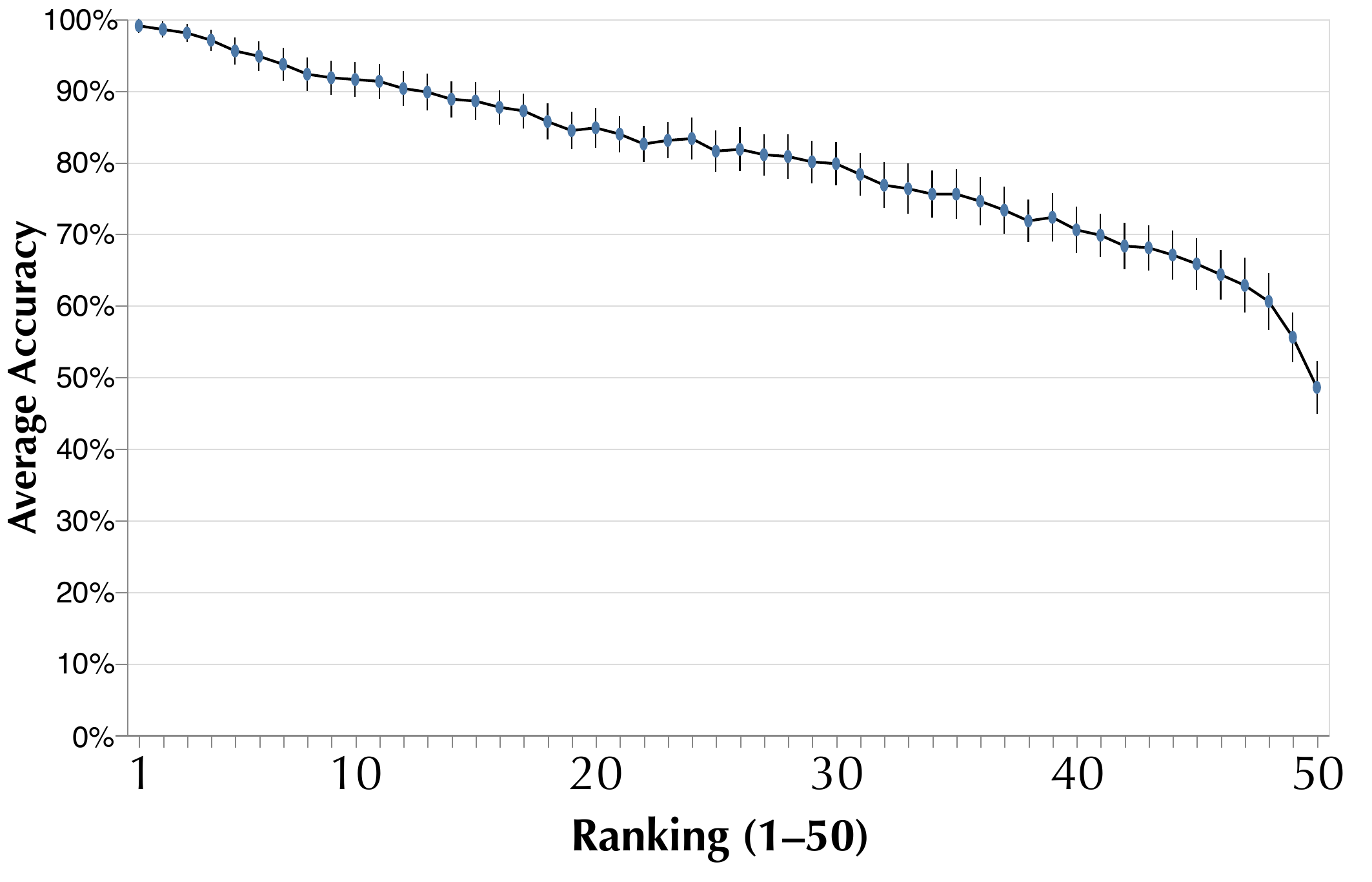} 
\vspace{0em}
\caption{\add{Cross-measure validation comparing the model-predicted ranking of redundant categorical palettes with human accuracy from Experiment 4. Our model predicts a ranking (1–50) for redundant palettes within each category number, where a higher rank indicates better expected performance. For validation, we randomly sampled 50 redundant palettes for each category number and repeated this process three times. The plot shows, for each ranking position, the mean human accuracy from our study along with 95\% confidence intervals, aggregated across all palettes assigned the same predicted rank. The monotonic decrease in performance provides preliminary evidence that CatPAW's ranking model aligns with empirical performance.}}
\label{fig:cross-validation}
\end{figure}

\add{Our model supports the unified design of color, shape, and redundant palettes by estimating their performance using three pairwise accuracy matrices. Given a set of candidate palettes, the model assigns a score to each palette and generates a ranking based on the predicted effectiveness. To evaluate the model’s ability to predict performance rankings for redundant palettes, we follow the approach used in Tseng et al. \cite{tseng2024shape} and perform a preliminary cross-measure validation comparing model predictions with ground-truth accuracy obtained in Experiment 4.}

\add{In Experiment 4, approximately 120 unique redundant palettes were tested for each category number, with each palette receiving 10 responses. For validation, we randomly selected 50 palettes per category number, repeated this sampling three times, and used our model to generate predicted rankings for each set. We then computed the actual mean human accuracy for palettes at each predicted rank. As shown in \autoref{fig:cross-validation}, there is a clear monotonic downward trend: palettes predicted to have lower within-category rankings consistently exhibited lower human accuracy across all category numbers. The correlation between predicted rank and mean human performance was strong ($r = 0.97, p < .0001$), indicating that the model closely aligns with empirical behavior. These findings offer preliminary evidence that our model can faithfully predict relative performance across redundant palettes. However, this validation focuses on ranked comparisons within our dataset; more extensive evaluation across broader datasets and task contexts remains an important direction for future work.}

\subsection{\add{Evaluation of \emph{CatPAW} against Designer and User-Select Categorical Palettes}}

\begin{figure}[htbp] 
\centering
\vspace{-0.2em}
\includegraphics[width=.45\textwidth]{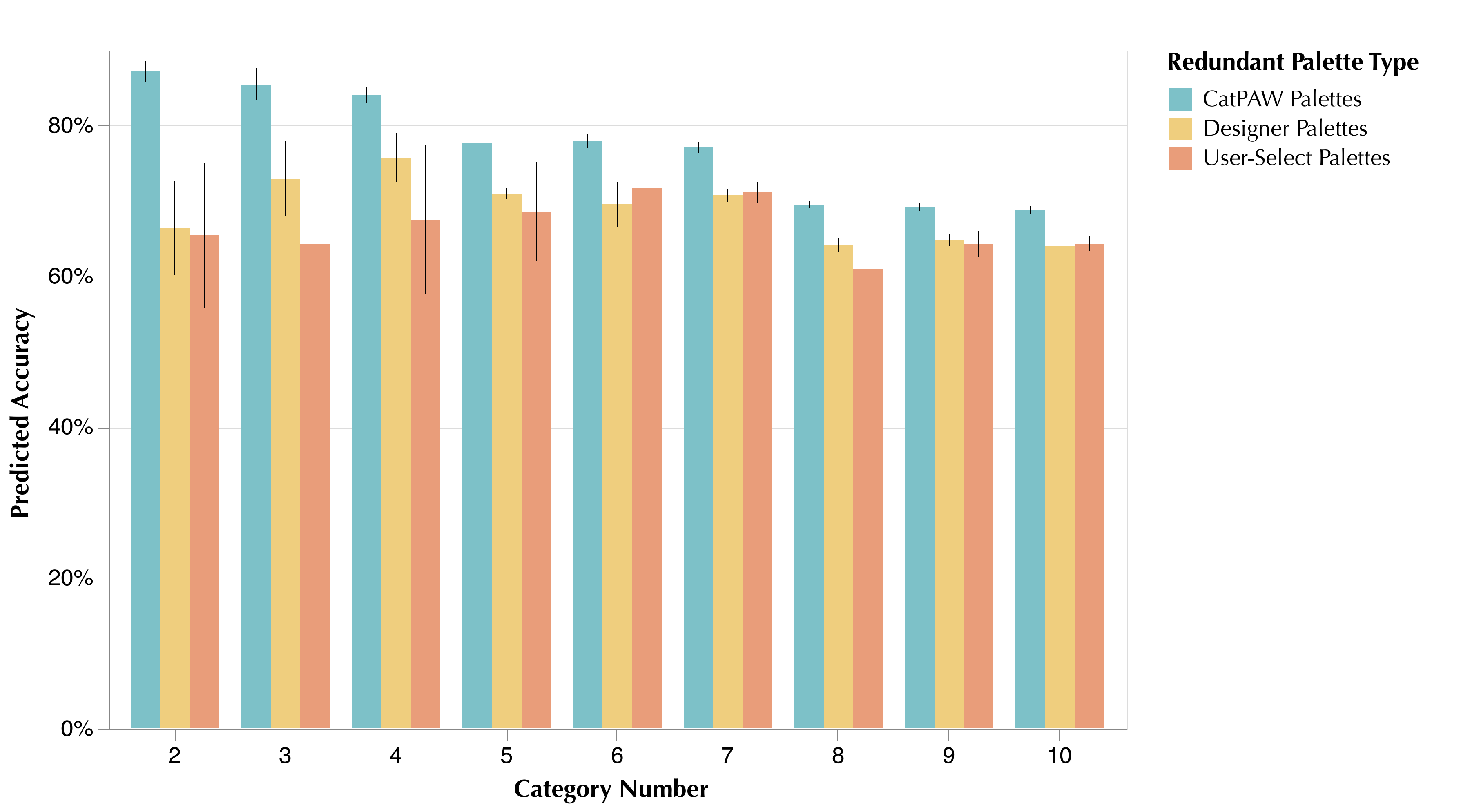} 
\vspace{0em}
\caption{\add{We compared the model-predicted performance of three groups of redundant categorical palettes across category numbers (2–10): \emph{CatPAW}-generated palettes, designer palettes, and user-selected palettes. Designer palettes were derived from widely used visualization systems (D3, Excel, MATLAB, and Tableau). We paired each system's color palette with its matching default shape palette and randomly combined these elements to create five redundant palettes per category number. For comparison, we sampled ten redundant palettes generated by CatPAW and ten palettes selected directly by users for each category size. Error bars show 95\% confidence intervals of the predicted accuracy.}}
\label{fig:palettes_evaluation}
\end{figure}

\add{
To the best of our knowledge, prior work offers no established guidance or design principles for constructing redundant color–shape palettes, despite their common use in visualization practice. As a result, there is no existing benchmark or standard which can be used to assess \emph{CatPAW}'s recommendations. 
To provide a preliminary assessment of \emph{CatPAW}'s palettes, we instead use the categorical encoding model to compare \emph{CatPAW}-generated palettes against two baselines: widely-used palettes from designer tools and palettes that users create on their own. Although \emph{CatPAW} is trained on this model, the goal of this analysis is not to claim a definitive performance advantage, but rather to characterize how \emph{CatPAW}’s data-driven recommendations differ from conventional design choices and to provide initial evidence of its potential utility.}

\begin{figure*}[htbp] 
\centering
\includegraphics[width=.9\textwidth]{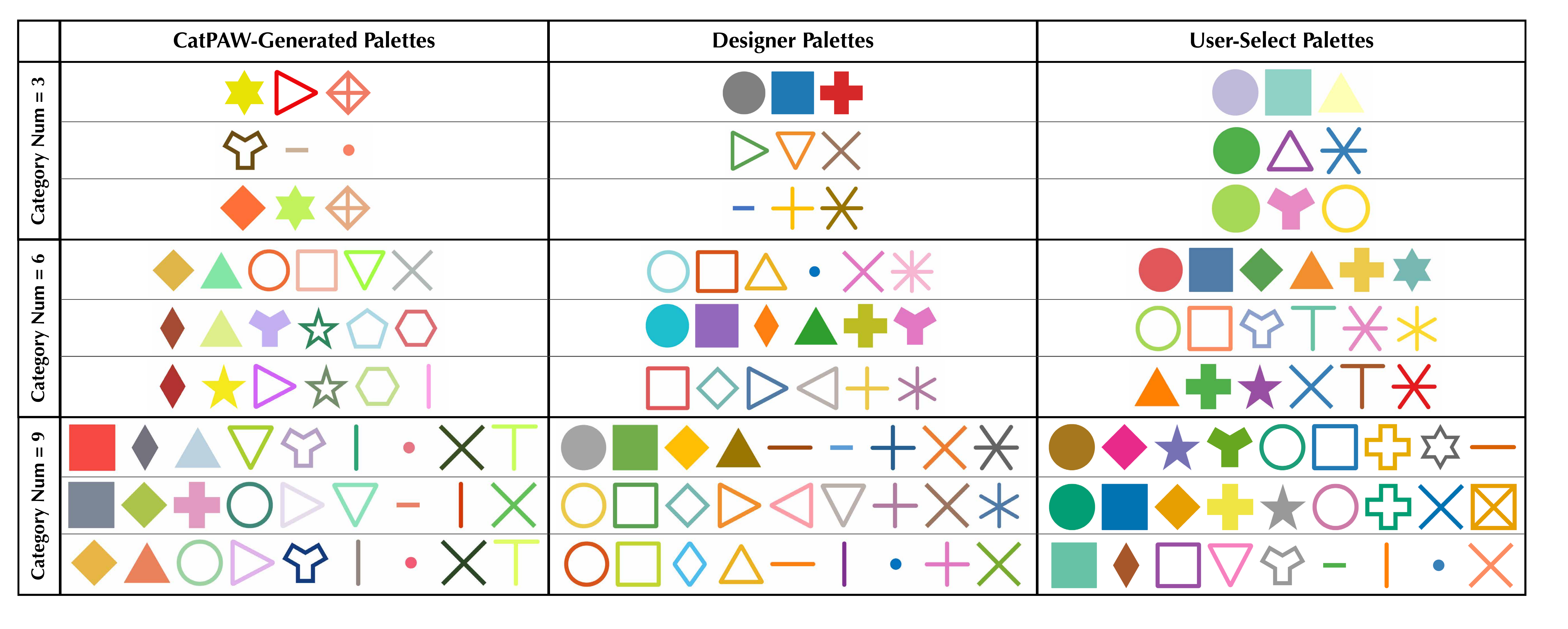} 
\vspace{-0.4em}
\caption{\add{Example redundant categorical palettes used in the model-based evaluation of palette quality. The figure shows three groups of palettes for category sizes 3, 6, and 9: \emph{CatPAW}-generated palettes, designer palettes, and user-selected palettes.}}
\label{fig:example_palettes_evaluation}
\end{figure*}

\add{
For the designer baseline, we selected one categorical color palette and one shape palette from each of D3, Excel, MATLAB, and Tableau. Because none of these systems provides guidance for how colors and shapes should be combined into redundant palettes, we randomly paired colors and shapes from each tool to construct five redundant palettes for each category number.}
\add{
To construct the user-selected palette baseline, we recruited ten participants from a local college
with self-reported experience creating visualizations for professional publication, who reflect representative target users for \emph{CatPAW}. Each participant was asked to design one redundant palette for each category number (2–10) that they believed would perform best for distinguishing categories. Participants had access to the pool of 39 shapes used in \emph{CatPAW} and were free to choose any colors they wished, either manually or by drawing directly from designer palettes. Examples of the palettes from three groups are shown in \autoref{fig:example_palettes_evaluation}.}
\add{
We then applied our categorical encoding model to generate predicted accuracy scores for every palette based on empirical data and computed the average accuracy for each palette group. As shown in \autoref{fig:palettes_evaluation}, \emph{CatPAW}-generated palettes (77.5\% mean predicted accuracy) outperformed both designer palettes (68.4\%) and user-selected palettes (66.2\%) across category sizes, demonstrating the potential value of data-driven palette construction. The performance gap is most pronounced for smaller category numbers (2–6), where \emph{CatPAW}’s predictions are consistently higher than both baselines. Designer palettes also outperformed user-selected palettes in these smaller category sizes, though the difference between them narrows as the number of categories increases. Overall, this analysis suggests that \emph{CatPAW} produces redundant palettes that are predicted to support correlation judgments more effectively than those created through traditional design tools or intuitive user choices. }

These findings provide preliminary evidence that empirical, model-driven palette generation may offer meaningful advantages.
\add{
However, this evaluation should be viewed as an early step toward understanding the comparative performance of \emph{CatPAW}. A more definitive assessment of \emph{CatPAW}’s utility will require a large-scale empirical study that directly measures task performance using \emph{CatPAW}-generated palettes, designer palettes, and user-created palettes with human participants. We see this as an important direction for future work.
}

\section{Discussion}

Despite being widely used in real-world applications, 
the role of redundancy in visualization remains contentious, with limited experimental data to demonstrate its utility 
or guide effective redundant encoding design. 
Our results 
confirm prior work demonstrating the benefits of redundant encoding for improving accuracy in visualization tasks~\cite{nothelfer2017redundant}.
Contrary to previous insights on smaller class numbers (2-3) that redundant encoding does not increase performance in multiclass scatterplots~\cite{gleicher2013perception}, we observed that combining color and shape information enhances the accuracy of correlation judgments. 
However, these benefits depend on the number of categories encoded.
We found that redundancy offered the greatest perceptual benefit for between 5 and 8 categories. 
For larger numbers of categories, the benefits of redundancy may begin to diminish as the task becomes either simply too difficult or the display too complex. This sensitivity to the number of categories may also explain the lack of benefit found in prior studies~\cite{gleicher2013perception}: 
for smaller category numbers, people may be better able to leverage the limited capacity of working memory to reason over categories~\cite{brady2013probabilistic, wang2025characterizing}, allowing them to attend to each category serially and readily process relevant information.

The efficacy of redundant designs depends not only on choosing effective color or shape palettes, but also in considering the mapping between color and shape. We found significant interactions between color and shape choices, which resonate with prior work on the limited separability of these visual channels~\cite{smart2019measuring, szafir2018modeling}.
For instance, 
unfilled and open shapes paired with high or low lightness color palettes tended to yield lower accuracy.
The interplay between color and shape for redundant encoding highlights the importance of a more holistic orientation to visualization design, considering not only visual channels in isolation but also how these channels (and encodings more generally) might be designed to complement one another. 
Our experiments unify past work that individually focused on color encodings~\cite{tseng2024revisiting} or shape encodings~\cite{tseng2024shape} to offer a more holistic approach to design.
Our research provides further evidence for these preliminary insights, offers specific guidance on how redundant encodings can mitigate this effect, and results in a universal tool %
to combine the benefits of both.

\subsection{Redundancy and Perceptual Processes}

To help bridge our findings with work in vision science and to understand the potential generalizability of our findings, we hypothesize several potential underlying perceptual mechanisms that may be at play in our study. 
By examining our results through the lens of these mechanisms, we offer a set of potential explanations that may explain both why redundant color-shape encodings improve accuracy in categorical perception---particularly within the 5–8 category range---and why specific shape-color pairings matter. 

The {cognitive load} associated with visualization viewing (i.e., the demand the task places on visual working memory) likely plays an \add{important} role in performance.
Past work has demonstrated that when visualizing fewer than six categories (termed as the \textit{subitizing limit}), viewers can efficiently process and compare categories using a single channel, such as color or shape, without overloading cognitive resources~\cite{wang2025characterizing}.
However, as the number of categories increases, the discriminability of categories within a single visual encoding channel diminishes, leading to significantly decreased performance.
Our results demonstrated that redundant encoding mitigates the reduced performance at the subitizing limit by providing a second perceptual cue, effectively reducing the cognitive effort required to distinguish between categories.
Furthermore, 
the gains associated with redundancy are highest 
near the subitizing limit~\cite{wang2025characterizing} (in this case, for 5-8 categories). This indicates
that the gains from redundant encoding are highest when less efficient perceptual mechanisms that place more demand on visual working memory are first employed. As a result, visual working memory may more efficiently process redundant encodings despite their increased complexity. 
Beyond approximately 8 categories, however, the benefits of redundancy may plateau or decline due to visual clutter or interference between channels, consistent with known theories in visual working memory~\cite{brady2013probabilistic}.

Second, we observed there are potential interaction effects between color and shape, indicating that not all pairings are equally effective.
For instance, unfilled or open shapes paired with very light or very dark color palettes led to lower accuracy.
This may be due to reduced perceptual salience: unfilled shapes have less colored area, which may weaken the effectiveness of color encodings \cite{smart2019measuring}.
Conversely, mixed-shape palettes (combining filled, unfilled, and open shapes) performed better across a wider range of color palettes, suggesting that variability in shape type enhances overall discriminability.
This supports the idea that effective redundant encoding relies on complementary channel strengths rather than simply doubling the same type of perceptual information~\cite{nothelfer2017redundant}.

Third, compared to hue, the significant impact of lightness ($L^*$) and chroma ($C^*$) magnitude in driving pairwise color accuracy suggests that they may be key factors in facilitating rapid visual segmentation between categories.
This is consistent with theories and evidence in early visual processing mechanisms that prioritize luminance and contrast over hue~\cite{ware2012information}. For example, evidence from magnetoencephalography (MEG) data shows that hue reaches a slower peak decoding, about 20 ms after luminance polarity, in visual systems~\cite{hermann2022temporal}.
Therefore, when combined with shape, these attributes may either reinforce or undermine the perceptual integrity of each mark together with the shape's visual properties (e.g., filled vs. outline). For example, lightness and chroma may drive initial segmentation, and then shape and hue help refine that segmentation.

We note that these hypotheses should be investigated in more controlled settings and using methods tailored to vision science-based inquiry. We offer them as a preliminary means for more broadly and systematically understanding redundancy and categorical encoding and for prompting deeper interdisciplinary inquiry.

\subsection{Design Implications}
We embed the results of our experiments into a system to support designers in creating effective categorical encodings. However, our results also raise several general guidelines for using redundancy effectively: 

\begin{itemize}
    \item \textbf{Prioritize Redundant Encodings for Moderate Category Numbers.}
    Utilizing redundant encodings unifying color and shape generally improves the accuracy of correlation estimation in multi-class scatterplots. This approach is particularly beneficial when dealing with a moderate number of categories (5-8 in this study).
    \item \textbf{Consider Color-Shape Pairing Strategies.}
    Designers should carefully consider how they pair colors and shapes, especially when creating scatterplots with more than 4 categories. Certain combinations can significantly impact performance. For example, our findings suggest that mixed-shape type palettes (i.e., those employing a combination of filled, unfilled, and open shapes) tend to work well with both light and dark color palettes, whereas palettes biased towards open shapes perform better with color palettes containing greater lightness variation.
    \item \textbf{Be Aware of the Influence of Color Attributes.}
    Color attributes such as lightness and perceptual distance influence the effectiveness of color-shape pairings. Avoid pairing unfilled or open shapes with color palettes that are skewed towards high or low lightness.
    \item \textbf{Balance Palette Complexity.}
    While redundant encoding can improve performance, it is important to avoid over-complicating a visualization by introducing unnecessary visual features or over-consuming limited visual channels. 
    \item \textbf{Prioritize Top-performing Color and Shape Pairs for Moderate Categories}
    As simply combining top-performing colors and shapes does not guarantee optimal results, designers should consider matching color and shape pairs with better pairwise performance.
\end{itemize}

\subsection{Limitations and Future Work}
While our study provides insight into how we can effectively design categorical visualization with redundant encodings, there are several directions for future work that will improve our understanding of redundant encoding and categorical visualization broadly. 
First, we 
assess only a subset of colors and shapes and only on a white background, which is among the most common background choices for modern visualizations. We employ a number of sampling strategies to maintain a broad yet tractable sampling of redundant palettes. 
Future work could explore a broader range of colors, shapes, category numbers, and background colors, as well as scatterplots with heavily overlapping configurations, to assess the generalizability of our findings. \add{While correlation is a common task for multivariate scatterplots \cite{sarikaya2017scatterplots},  understanding performance across other tasks, such as outlier detection or cluster identification, and visualization idioms like bar charts and icon arrays, is also important future work to assess the generalizability of our findings.}
Such work should also investigate associations between specific shape and color attributes, such as hue variation or shape features, and performance to 
provide a more nuanced understanding of color-shape interactions.

Our study focuses purely on abstract palette design. Other factors beyond simply color, shape, and category number may influence palette effectiveness. For example, our study did not explicitly account for participant preferences or aesthetic factors in colors and shapes,
which may also impact their responses~\cite{schloss2013object}. We also focused on palettes for abstract, neutral values; however, both shapes and colors can carry semantic meaning that may influence their effectiveness in isolation as well as in redundant pairs \cite{schloss2018mapping}. Future research could investigate how these factors influence the effectiveness of redundant encodings and categorical palette design. Further, redundancy can also be applied with ordinal and numeric data. Our findings offer hypotheses about general redundant encodings, but future work should confirm whether our observations generalize to quantitative redundant encodings. 

\emph{CatPAW} represents a step towards automated palette generation by considering redundant encodings. 
It does not explore using colors and shapes to encode different data dimensions simultaneously nor can it currently adapt to new shapes. Future work could focus on refining the underlying model, incorporating more sophisticated optimization techniques, and accounting for accessibility to create even more effective and user-friendly design tools. \add{Further, future work should evaluate CatPAW's performance across a broader range of tasks, factors such as aesthetic preference, and in more formal human subjects experiments. } As CatPAW is an extensible open-source tool, future results, such as new shapes or colors, can be readily integrated into the underlying data model to generate a wider range of improved predictions.

\section{Conclusion}
We investigated the effectiveness of redundant encoding by systematically evaluating how color and shape pairings influence categorical perception in data visualization. Our findings demonstrated that while redundant encodings can enhance task performance in multiclass scatterplots, their benefits are most evident within a certain range of category numbers (5-8). Furthermore, we highlight the importance of combining colors and shapes and provide a set of guidelines for designing effective redundant categorical palettes. To translate these insights into an actionable framework, we developed a data-driven model informed by experimental results and implemented it in a web-based tool that enables designers to create categorical palettes with flexible user preferences. We hope this work will encourage future research toward developing comprehensive strategies for categorical data visualization.

\begin{acks}
We thank the reviewers for their insightful comments. This work was supported by the National Science Foundation under grant No.2127309 to the Computing Research Association for the CIFellows project and NSF IIS-2320920.
\end{acks}

\newpage
\bibliographystyle{ACM-Reference-Format}
\bibliography{main}

\end{document}